\documentclass[useAMS,usenatbib]{mn2e}
\usepackage{astrophys,graphicx}
\title[ULXs in NGC\,2276]{NGC\,2276: a remarkable galaxy with a large number of ULXs}
\author[A.~Wolter et al.] {Anna~Wolter,$^{1}$\thanks{E-mail: anna.wolter@brera.inaf.it}
Paolo Esposito,$^{2,3}$
Michela Mapelli,$^{4}$  
Fabio Pizzolato$^{5}$ and
\newauthor
Emanuele Ripamonti$^{6}$
 \smallskip\\
$^1$Osservatorio Astronomico di Brera, INAF, via Brera 28, I-20121 Milano, Italy
\\
$^2$IASF--Milano, INAF, via E. Bassini 15, I-20133 Milano, Italy\\
$^3$Harvard--Smithsonian Center for Astrophysics, 60 Garden Street,
Cambridge, MA 02138, USA\\
$^4$Osservatorio Astronomico di Padova, INAF, vicolo dell'Osservatorio 5, I-35122 Padova, Italy\\
$^5$Dipartimento di Fisica, Universit\`a degli Studi di Milano, via G. Celoria 16, I-20133 Milano, Italy\\
$^6$Dipartimento di Fisica e Astronomia `Galileo Galilei', Universit\`a degli Studi di Padova,\\ vicolo dell'Osservatorio 3, I-35122 Padova, Italy
}

\date{Accepted 2015 January 7.  Received 2015 January 7; in original form 2014 September 6} \pagerange{\pageref{firstpage}--\pageref{lastpage}} \pubyear{2015}

\def\LaTeX{L\kern-.36em\raise.3ex\hbox{a}\kern-.15em
    T\kern-.1667em\lower.7ex\hbox{E}\kern-.125emX}

\def\xmm {\emph{XMM--Newton}}
\def\cxo {\emph{Chandra}}

\def\rst {\emph{ROSAT}}

\def\flux {\mbox{erg cm$^{-2}$ s$^{-1}$}}
\def\lum {\mbox{erg s$^{-1}$}}
\def\nh {$N_{\rm H}$}
\def\lesssim{\ \raise -2.truept\hbox{\rlap{\hbox{$\sim$}}\raise5.truept  
\hbox{$<$}\ }}                        % minore o circa uguale

\begin{document}

\label{firstpage}
\maketitle
\begin{abstract}
The starbusting, nearby ($D = 32.9$~Mpc) spiral (Sc) galaxy NGC\,2276 belongs to the sparse group dominated by the elliptical galaxy NGC\,2300. NGC\,2276 is a remarkable galaxy, as it displays a disturbed morphology at many wavelengths. This is possibly due to gravitational interaction with the central elliptical galaxy of the group. Previous \rst\ and \xmm\ observations resulted in the detection of extended hot gas emission and of a single very bright ($\sim$$10^{41}$~\lum) ultraluminous X-ray source (ULX) candidate. Here we report on a study of the X-ray sources of NGC\,2276 based on \cxo\ data taken in 2004. \cxo\ was able to resolve 16 sources, 8 of which are ULXs, and to reveal that the previous ULX candidate is actually composed of a few distinct objects. We construct the luminosity function of NGC\,2276, which can be interpreted as dominated by high mass X-ray binaries, and estimate the star formation rate (SFR) to be $\sim$5--15~M$_{\sun}$~yr$^{-1}$, consistent with the values derived from optical and infrared observations. By means of numerical simulations, we show that both ram pressure and viscous transfer effects are necessary to produce the distorted morphology and the high SFR observed in NGC\,2276, while tidal interaction have a marginal effect.
\end{abstract}

\section{Introduction}
\label{Intro}

NGC\,2276 is an Sc galaxy belonging to the poor group of galaxies NGC\,2300, which is composed of four or five galaxies\footnote{Apart from NGC\,2300 and NGC\,2276, the other members are NGC\,2268 and IC\,455 \citep{hg82}. UCG\,03670, at 16$'$ ($\sim$0.15~Mpc) from NGC\,2300, is likely another group member.} and dominated by the namesake elliptical galaxy. Here and throughout the paper we assume a distance\footnote{From the NASA/IPAC Extragalactic Database (NED)  \mbox{http://ned.ipac.caltech.edu/} and adopting $H_0 = 73$~km~s$^{-1}$~Mpc$^{-1}$} to NGC\,2276 of 32.9~Mpc and rescale published values to this distance when necessary. The NGC\,2300 group is the first in which an X-ray emitting intragroup medium (IGM) was observed \citep{mulchaey93}. Observations with \rst\ in fact revealed an extended ($\simeq$0.2~Mpc), unusually dense ($\simeq 5.3\times 10^{-4}$~cm$^{-3}$), hot ($\simeq$0.9~keV) and relatively metal poor ($\simeq$0.06~Z$_{\sun}$) intragroup gas halo \citep{mulchaey93,davis96}. NGC\,2276 is undergoing starburst activity. A number of H\textsc{ii} regions were identified in the galaxy, spread throughout its volume \citep{hodge83,davis97}, and many supernovae have been observed in the last fifty years (e.g. \citealt*{iskudaryan67,barbon89}, \citealt{treffers93}, \citealt*{dimai05}). The star formation rate (SFR) was estimated by \citet{kennicutt83} to be 9.5~M$_{\sun}$~yr$^{-1}$ based on the high H$\alpha$ luminosity, and values between 5 and 15~M$_{\sun}$~yr$^{-1}$ can be found in the literature from different measurements (mostly H$\alpha$ and infrared observations; e.g. \citealt*{kennicutt94}, \citealt{sanders03,james04short}).

The asymmetric shape of NGC\,2276, with a `bow-shock-like' structure along its western edge and a `tail' of gas extending towards east, and the unusually intense star formation, especially along the western edge, (the source falls in the upper range of the distribution of star-forming galaxies; see e.g. \citealt{noeske07} for the scatter-plots of galaxies in the Aegis field) have attracted considerable attention but, while it is quite clear that these features are due to interaction with the environment, no consensus on the dominant physical process involved has yet been reached.

Since NGC\,2300 and NGC\,2276 are relatively close to each other (7$'$, about 70~kpc), a prime candidate is the tidal interaction between the two galaxies \citep{gruendl93,davis96,davis97,elmegreen91}.
Other processes involved in shaping NGC\,2276 are the viscous and the ram-pressure stripping of the galactic gas in the IGM \citep*{nulsen82,rasmussen06}. \citet{rasmussen06} questioned the importance of the tidal interactions and argued that turbulent viscous and ram-pressure stripping, ram pressure compression of the disc gas and starburst outflows are more likely the main elements influencing the shape and SFR of NGC\,2276.

Using an \xmm\ observation carried out in 2001, \citet{davis04} identified a bright X-ray source (XMMU\,J072649.2+854555) at the western edge of the galaxy with an ultraluminous X-ray source (ULX: non-nuclear sources with X-ray luminosities in excess of $L_{\mathrm{X}} = 10^{39}$\lum\ ; see e.g. \citealt{feng11} for a review) candidate. The X-ray luminosity of this object in the 0.5--10~keV band was $L_{\mathrm{X}}^{0.5-10\,\mathrm{keV}} = 1.1\times10^{41}d^2_{45.7}=5.7\times10^{40}d^2_{32.9}$~\lum\ (where $d_N$ stands for the distance in units of $N$~Mpc), making it one of the most luminous known in its class (e.g. \citealt{sutton12}). They also observed that the nuclear source, undetected in the \xmm\ data, was much dimmer (by a factor of $\sim$10 at least) than the \rst/HRI nuclear source seen by \citet{davis97} eight years earlier at $L_{\mathrm{X}}^{0.5-2\,\mathrm{keV}}\sim 2\times10^{40}d^2_{45.7}=10^{40}d^2_{32.9}$~\lum. 

ULXs are mainly found in galaxies with high SFRs, and indeed most might be assumed to be the extension of the High Mass X-ray Binaries (HMBX) population to high X-ray luminosities. For instance, in The Antennae \citep{zf02}, in the Cartwheel galaxy \citep{wolter04}, in the NGC\,2207/IC\,2163 pair \citep{mineo13} and in NGC\,4088 \citep{mezcua14} the X-ray luminosity function follows the ``universal'' law found by \citet{grimm03} for HMXBs when scaled for the SFR.
This assumption, linked to the proposed influence of low metallicity on the evolution of massive stars \citep[see][and references therein]{mapelli10} leads to expect not one, but a large number of ULXs in NGC\,2276. 
%The finding of a single ULX in NGC\,2276 was uncomfortable within the scenarios which assume that ULXs are the brightest tail of the high mass X-ray binary (HMXB) population. {\bf ref only for .Lx < 1-5e+40 erg/s?}  
Using the correlations by \citet{mapelli10} and assuming a metallicity $Z=0.22$~Z$_{\sun}$ and a SFR of $10$~M$_{\sun}$~yr$^{-1}$, we expect about 10 ULXs. The metallicity is derived applying the \citet{pilyugin05} calibration on the de-reddened [OII\,$\lambda$ 3727]/H${\beta}$ and [OIII\,$\lambda$5007]/H${\beta}$ ratios measured by \citet{kennicutt92} on an integrated spectrum of NGC2276. Indeed, a high-spatial-resolution observation of NGC\,2276 taken with \cxo\ (Obs.ID 4968, PI: Rasmussen) revealed that the bright ULX candidate is actually composed of a few different sources \citep{wolter11}, and showed a total of 8--9 sources bright enough ($L_{\mathrm{X}} > 10^{39}$~\lum) during the observation to be classified as ULXs \citep{wolter11,liu11}.  

The \cxo\ data allowed \citet{rasmussen06} to resolve details of the X-ray morphology of NGC\,2276, which, as seen at other wavelengths, is characterised by a shock-like feature along the western edge and a low surface brightness tail extending to the east. Spatially resolved spectroscopy shows that the data are consistent with the disc gas being pressurized at the leading western edge of NGC\,2276, due to the galaxy moving supersonically through the IGM, at a velocity of $\approx$900~km~s$^{-1}$. \citet{rasmussen06} estimated that the diffuse hot gas in the disc of NGC\,2276 has a temperature of $kT \sim 0.3$~keV and an X-ray luminosity of $L_{\mathrm{X}}^{(0.3-2.0\,\mathrm{keV})} = 1.9\times 10^{40}d^2_{36.8}=1.5\times10^{40}d^2_{32.9}$~\lum, with an inferred residual component of unresolved X-ray binaries of about 10--15 per cent of the total flux. The IGM, as observed by \cxo, has a profile consistent with the previous \rst\ measurements \citep{davis96}, a temperature of $kT\approx0.9$~keV and $Z=0.17$~Z$_{\sun}$ \citep{rasmussen06}. This metallicity is well matched, given also the different method used, to the one used above.

In \citet{wolter11}, we reported on the several new ULXs observed by \cxo\ in NGC\,2276 and on the fact that the bright ULX candidate XMMU\,J072649.2+854555 actually consists of a number of distinct sources (see also \citealt{sutton12}). In this work, we give more details on the point sources, presenting the whole \cxo\ data analysis, investigate the main mechanisms at work in shaping the galaxy by means of numerical simulations, and discuss their influence on the X-ray source population.

\section{Analysis of \emph{CHANDRA} data}\label{anal}

The \cxo\ observation (Obs. ID 4968, see \citealt{rasmussen06} for further details) was performed on 2004 June 23 with the Advanced CCD Imaging Spectrometer (ACIS; \citealt{garmire03}), and lasted about 45.8~ks. The ACIS was operated in the standard timed exposure full-frame mode, with the `very faint' telemetry format, and NGC\,2276 was positioned on the back-illuminated chip S3 (CCD7). The data were reprocessed with the \cxo\ Interactive Analysis of Observations software package (\textsc{ciao}, version 4.6). \footnote{http://cxc.harvard.edu/ciao/} We removed the pixel randomization, since it deteriorates the spatial resolution, but, other than that, we followed standard data reduction and analysis procedures. 

The first step of our analysis was to run the \textsc{ciao} wavelet source-detection algorithm with \textsc{sigthresh}\footnote{See http://cxc.harvard.edu/ciao/ahelp/wavdetect.html for the parameters of \textsc{wavdetect}.} set to $10^{-6}$, so that we expect at most one spurious source in the area under examination given that we consider only a single CCD. The routine found 41 sources in CCD7, of which 16 in the galaxy area, defined by $D25 = 177\farcs1$ \citep{devaucouleurs91} with a signal-to-noise ratio higher than 4.
\begin{figure*}
\centering
%\resizebox{\hsize}{!}{\includegraphics[]{fig1opt+xray.ps}}
\resizebox{\hsize}{!}{\includegraphics[]{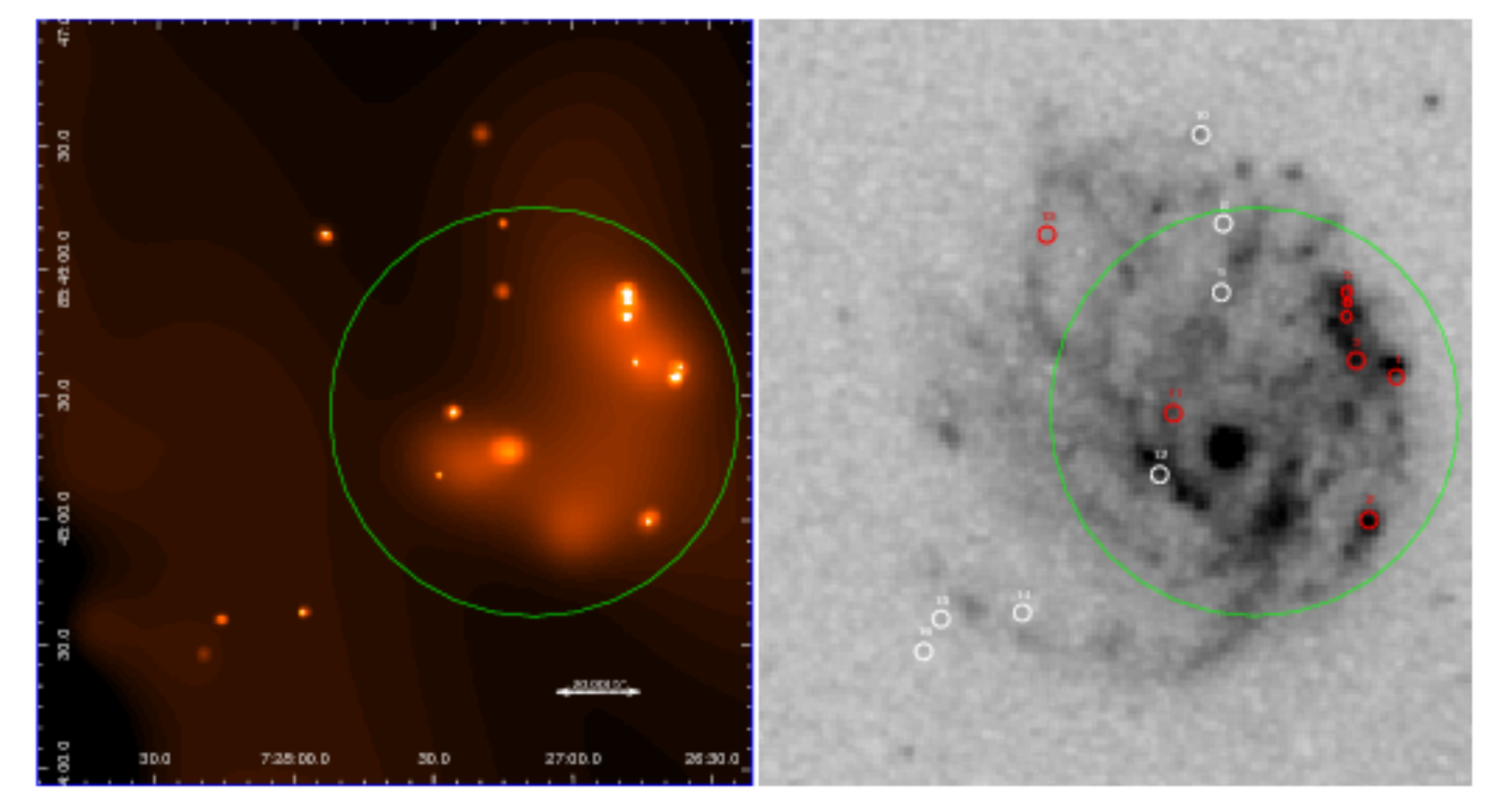}}
\caption{\label{images}{\it Left:} Smoothed image (\textsc{csmooth}) from \cxo\ data in the 0.5--7 keV band. The green circle (radius=50$^{\prime\prime}$) indicates the region used to extract the diffuse component spectrum (associated to S7, see text for details) {\it Right:} The X-ray sources' positions are over-plotted on the optical image of NGC\,2276 taken from the  Digitized Sky Survey (DSS), see http://archive.stsci.edu/dss/index.html.  Open circles and numbers identify detected point sources; red sources are ULXs. White sources have $L_{\mathrm{X}}^{0.5-10} \leq 10^{39}$\lum.
}
\end{figure*}
\begin{table}
\centering
\caption{Source detected in NGC\,2276. Positions from the detection algorithm and net counts from the extraction regions of 2$^{\prime\prime}$ radius (see text).
No astrometric correction has been applied and therefore the positional uncertainty is the standard \cxo\ one of $\Delta \leq 0\farcs6$ (90\% c.l.). }
\label{src}
\begin{tabular}{@{}lccr}
\hline
Number & RA (J2000) & Dec (J2000) & Net Counts \\ 
   & & & (0.3--5~keV) \\
\hline
 S1 & 07$^{\rm h}$26$^{\rm m}$37$\fs$4 & +85$\degr$45$'$34$\farcs$5 & $417.4\pm20.4$ \\ 
 S2 & 07$^{\rm h}$26$^{\rm m}$43$\fs$5 & +85$\degr$45$'$00$\farcs$1  &    $46.2 \pm    6.9$ \\
 S3 & 07$^{\rm h}$26$^{\rm m}$46$\fs$1 & +85$\degr$45$'$38$\farcs$5  &    $43.2 \pm    6.6$ \\
 S4 & 07$^{\rm h}$26$^{\rm m}$47$\fs$9 & +85$\degr$45$'$52$\farcs$3  &   $175.8 \pm   13.3$ \\
 S5 & 07$^{\rm h}$26$^{\rm m}$48$\fs$1 & +85$\degr$45$'$54$\farcs$9  &   $338.6 \pm   18.4$ \\
 S6 & 07$^{\rm h}$26$^{\rm m}$48$\fs$2 & +85$\degr$45$'$49$\farcs$0  &   $104.7 \pm   10.2$ \\
 S7$^{a}$ & 07$^{\rm h}$27$^{\rm m}$13$\fs$0 & +85$\degr$45$'$16$\farcs$0  &  $40.3  \pm  6.4 $ \\
 S8 & 07$^{\rm h}$27$^{\rm m}$14$\fs$9 & +85$\degr$46$'$11$\farcs$5  &    $14.2 \pm    3.9$ \\
 S9 & 07$^{\rm h}$27$^{\rm m}$15$\fs$4 & +85$\degr$45$'$54$\farcs$9  &     $8.2 \pm    3.0$ \\
S10 & 07$^{\rm h}$27$^{\rm m}$19$\fs$7 & +85$\degr$46$'$32$\farcs$9  &    $19.2 \pm    4.5$ \\
S11 & 07$^{\rm h}$27$^{\rm m}$25$\fs$8 & +85$\degr$45$'$25$\farcs$9  &    $68.2 \pm    8.3$ \\
S12 & 07$^{\rm h}$27$^{\rm m}$28$\fs$7 & +85$\degr$45$'$11$\farcs$1  &    $21.3 \pm    4.7$ \\
S13 & 07$^{\rm h}$27$^{\rm m}$53$\fs$3 & +85$\degr$46$'$08$\farcs$8  &    $53.2 \pm    7.3$ \\
S14 & 07$^{\rm h}$27$^{\rm m}$58$\fs$4 & +85$\degr$44$'$37$\farcs$9  &    $19.3 \pm    4.5$ \\
S15 & 07$^{\rm h}$28$^{\rm m}$16$\fs$0 & +85$\degr$44$'$36$\farcs$4  &    $29.2 \pm    5.5$ \\
S16 & 07$^{\rm h}$28$^{\rm m}$19$\fs$7 & +85$\degr$44$'$28$\farcs$5  &    $13.3 \pm    3.7$ \\
\hline
\end{tabular}
\begin{list}{}{}
\item[$^{a}$]Source S7 is the center of the galaxy: for subsequent analysis counts have been extracted in a circle of 50$^{\prime\prime}$ radius, excluding all detected sources, resulting in a total of $1413 \pm  67$ net counts.
\end{list}
\end{table}

Source positions (in RA order) and statistics of the 16 sources are given in Table\,\ref{src}, while in Fig.\,\ref{images} they are shown in the X-ray image and over-plotted on an optical image
%\footnote{From the Digitized Sky Survey (DSS), see http://archive.stsci.edu/dss/index.html} 
of the galaxy. Source counts were extracted from circular regions centered at the positions found by the source-detection routine and with radius 2$''$. The sources are all at $< 2^{\prime}$ from the aimpoint, therefore the 2$''$ circle is large enough to include more than 90\% of all counts \footnote{See e.g. fig.\,4.7 in http://cxc.harvard.edu/proposer/POG/html/chap4.html} and is a very good match to the ellipse from the detection algorithm. We expect a very small contribution in the area from both the instrumental and the diffuse background. The dominant source of error is statistic. 
For source S7, which corresponds to the nuclear region, we extracted counts in a circle of 50$''$ radius (excluding all the detected sources in that area) for consistency with the \citet{rasmussen06} analysis (see Section 4). The sources' positions are plotted in Fig.\,\ref{images} and the extraction area for S7 is indicated as the green circle. Instrumental and cosmic background was estimated from circular regions outside of the galaxy, but as close as possible to it, and excluding detected sources. We did not attempt to subtract the diffuse gas contribution, which is, at any rate, negligible in the extraction area of the point sources (less than 1 count per source). A first indication of the sources' spectral shape can be evinced from Fig.\,\ref{truecolor} in which the ``true'' color image of the X-ray emission has been constructed. The absorbed or flat sources appear as blue/green in this image.For a more formal analysis, we extracted individual spectra for the sources with more than a hundred net counts (S1, S4, S5, and S6, plus the extended region around S7). For the other sources (S2, S3, S8, S9, S10, S11, S12, S13, S14, S15, S16), since the paucity of counts hinders individual fits, we extracted a cumulative spectrum, which contains $336\pm20$ net counts. All the spectral analysis was performed with the \textsc{xspec} (v. 12.8.0) fitting package \citep{arnaud96}. Results and analysis of the spectra are described in the next Sections.
\section{Point sources and ULXs}
\label{pntsrcs}

\begin{table*}
\begin{minipage}{14.3cm}
\centering
\caption{Results of spectral fit for the 4 ULXs with more than 100 net counts and for the sum of the fainter sources.
Unabsorbed fluxes and luminosities are in the 0.5--10~keV energy band. 
}
\label{fit}
\begin{tabular}{@{}lccrrrr}
\hline                                    
Name &   $\Gamma$  & \nh  & $F_{\mathrm{X}}^{(0.5-10\,\mathrm{keV})}$  &    $L_{\mathrm{X}}^{(0.5-10\,\mathrm{keV})}$ &   $L_{\mathrm{X}}^{(2-10\,\mathrm{keV})}$ & $\chi^2_\nu$~(dof)\\
     &   & ($10^{21}$~cm$^{-2}$)  & ($10^{-14}$~\flux) & ($10^{39}$~\lum)  & ($10^{39}$~\lum) \\
\hline
S1 & $1.96\pm0.34$ & $4.1\pm2.0$ & $14.45\pm0.71$ & 18.7 & 10.29 & 0.92 (17)\\
S4 & $1.48^{+0.49}_{-0.25}$ & $<25.0$ &  $6.78\pm0.51$ & 8.73 & 6.39 & 1.47 (5)\\
S5 & $1.77\pm0.40$ & $3.1\pm2.0$ & $11.10\pm0.60$ & 14.3 & 8.90 & 1.54 (13)\\
%S6$^a$ & $> 3.2$ & $6.0\pm4.0$ & $< 1.6e-11 --6.21 in 2-10 $ & $< 2.42e42$ & $<9.3$ & 1.50 (1)\\  <-- questi valori sono calcolati con Gamma  best fit = 6.01
S6$^a$ & $> 3.2$ & $6.0\pm4.0$ & $2.60\pm0.25$ & 3.36 & 2.86 & 1.50 (1)\\
SUM & $1.93\pm0.35$ & $1.7\pm0.9$ & $6.00\pm0.37$ & 7.74 & 4.41 & 1.11 (18) \\ 
\hline
\end{tabular}
\begin{list}{}{}
\item[$^{a}$] Flux and luminosities were computed with  $\Gamma = 3.2$ and the fitted value for \nh = $25.3 \times 10^{21}$~cm$^{-2}$.
\end{list}
\end{minipage}
\end{table*}

\begin{figure}
%\hbox{\includegraphics[width=8.3cm]{N2276threecolorHR3.ps}}
\hbox{\includegraphics[width=8.3cm]{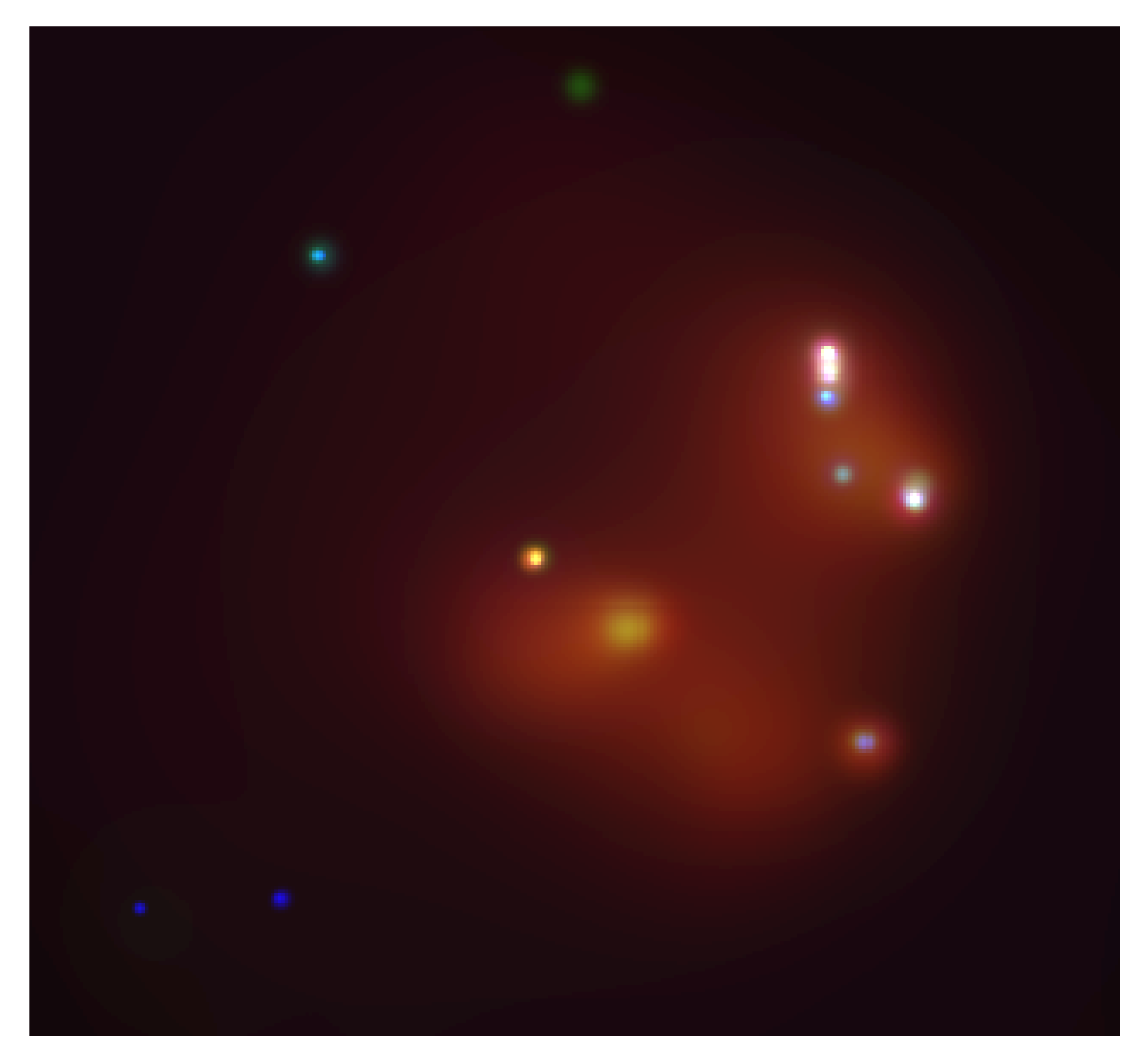}}
\caption{\label{truecolor}Three color image of the \cxo\ data. Smoothed with \textsc{csmooth} from \textsc{ciao} in the 0.3--0.9 keV, 0.9--1.5 keV and 1.5--8 keV bands for red, blue and green respectively. Many point sources appear blu/green indicating excess absorption and/or hard spectra.  }
\end{figure}

The distribution of the detected sources follows roughly the position of the arms of NGC\,2276 and the zones of high activity in the other bands. Contamination from background sources may be estimated according to the $\log N$--$\log S$ distribution of \citet{hasinger93}. At the detection limit of $F_{\mathrm{X}}^{(0.5-2\,\mathrm{keV})} = 7 \times 10^{-16}$~\flux\ (which corresponds to $L_{\mathrm{X}} \approx 10^{38}$~\lum\ at the distance of NGC\,2276) we expect 1.4 sources by chance in the total area covered by the galaxy. Therefore, at least one of the fainter sources is probably a background source.

\begin{figure}
\centering
%\resizebox{0.95\hsize}{!}{\includegraphics[angle=-90]{s1sp.ps}}
%\resizebox{0.95\hsize}{!}{\includegraphics[angle=-90]{s4sp.ps}}
%\resizebox{0.95\hsize}{!}{\includegraphics[angle=-90]{s5sp.ps}}
%\resizebox{0.95\hsize}{!}{\includegraphics[angle=-90]{s6sp.ps}}
\resizebox{0.95\hsize}{!}{\includegraphics[angle=-90]{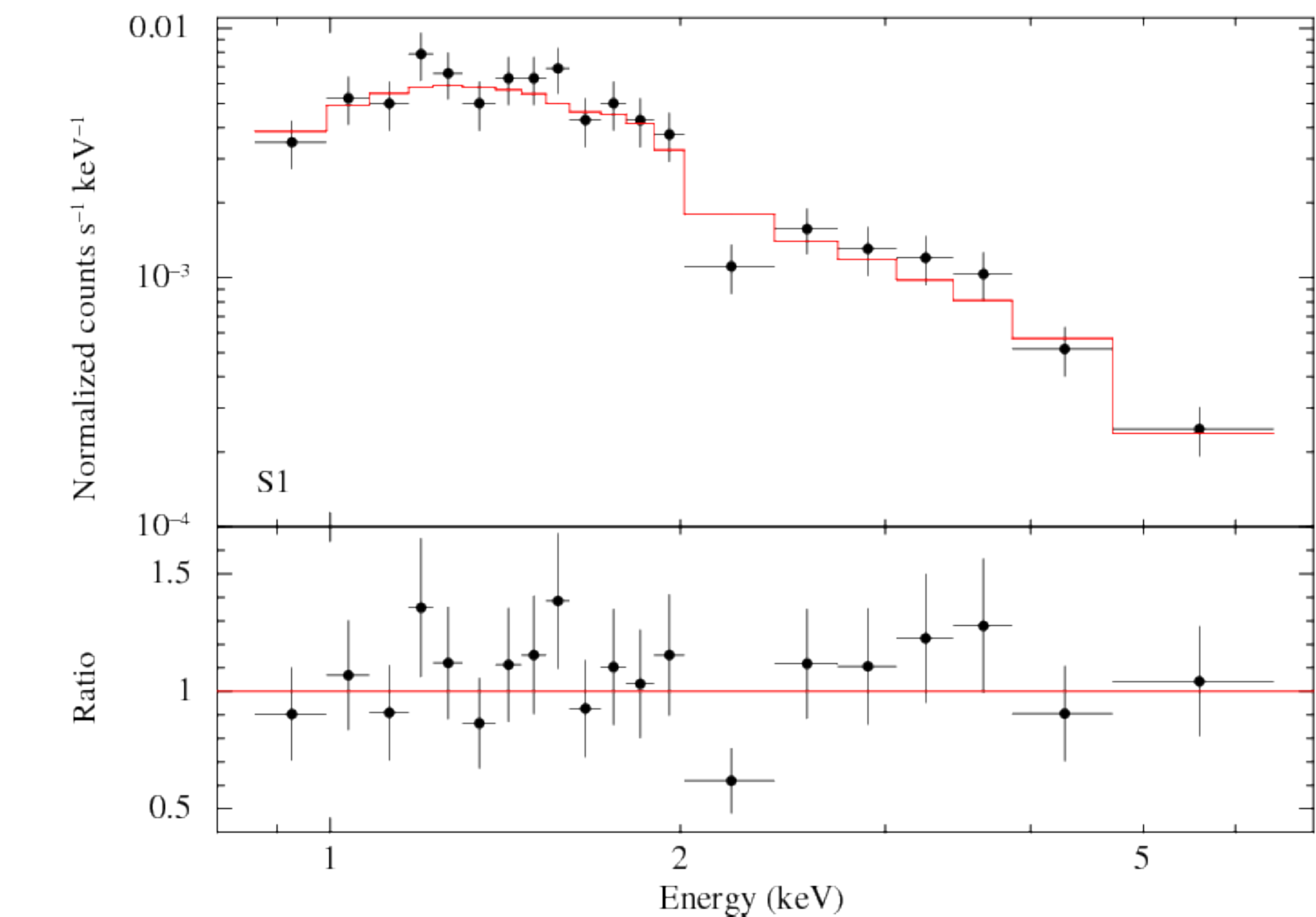}}
\resizebox{0.95\hsize}{!}{\includegraphics[angle=-90]{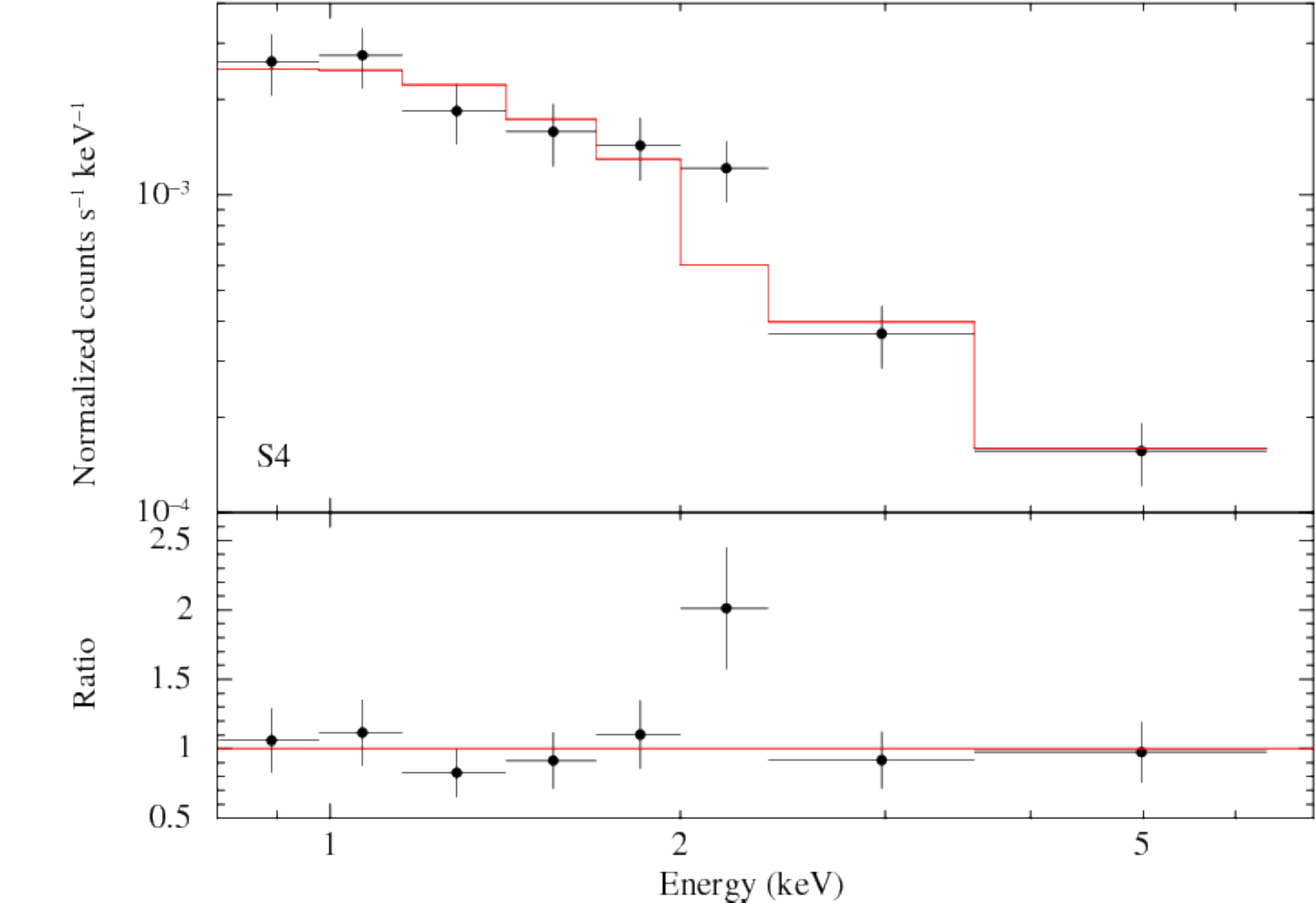}}
\resizebox{0.95\hsize}{!}{\includegraphics[angle=-90]{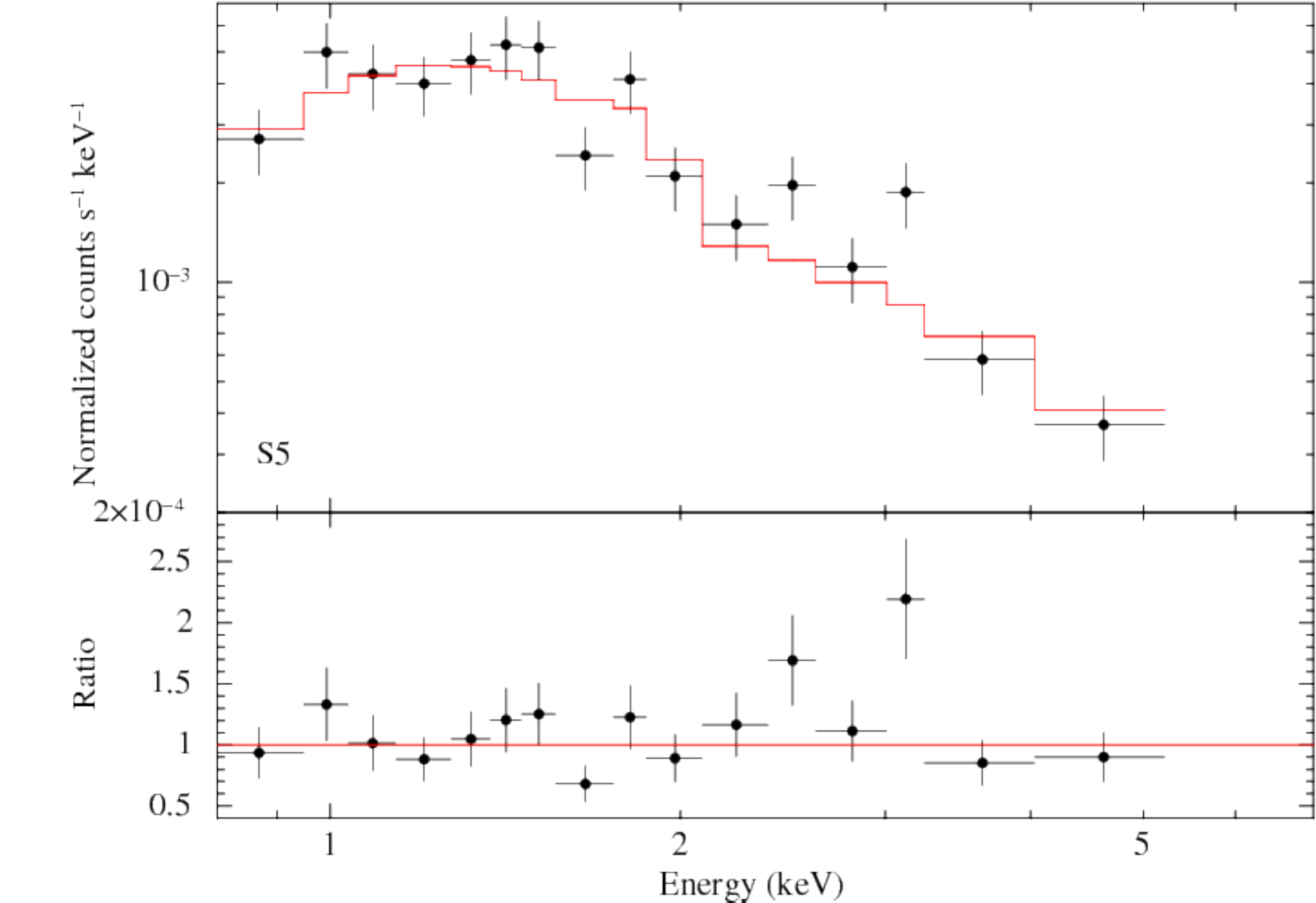}}
\resizebox{0.95\hsize}{!}{\includegraphics[angle=-90]{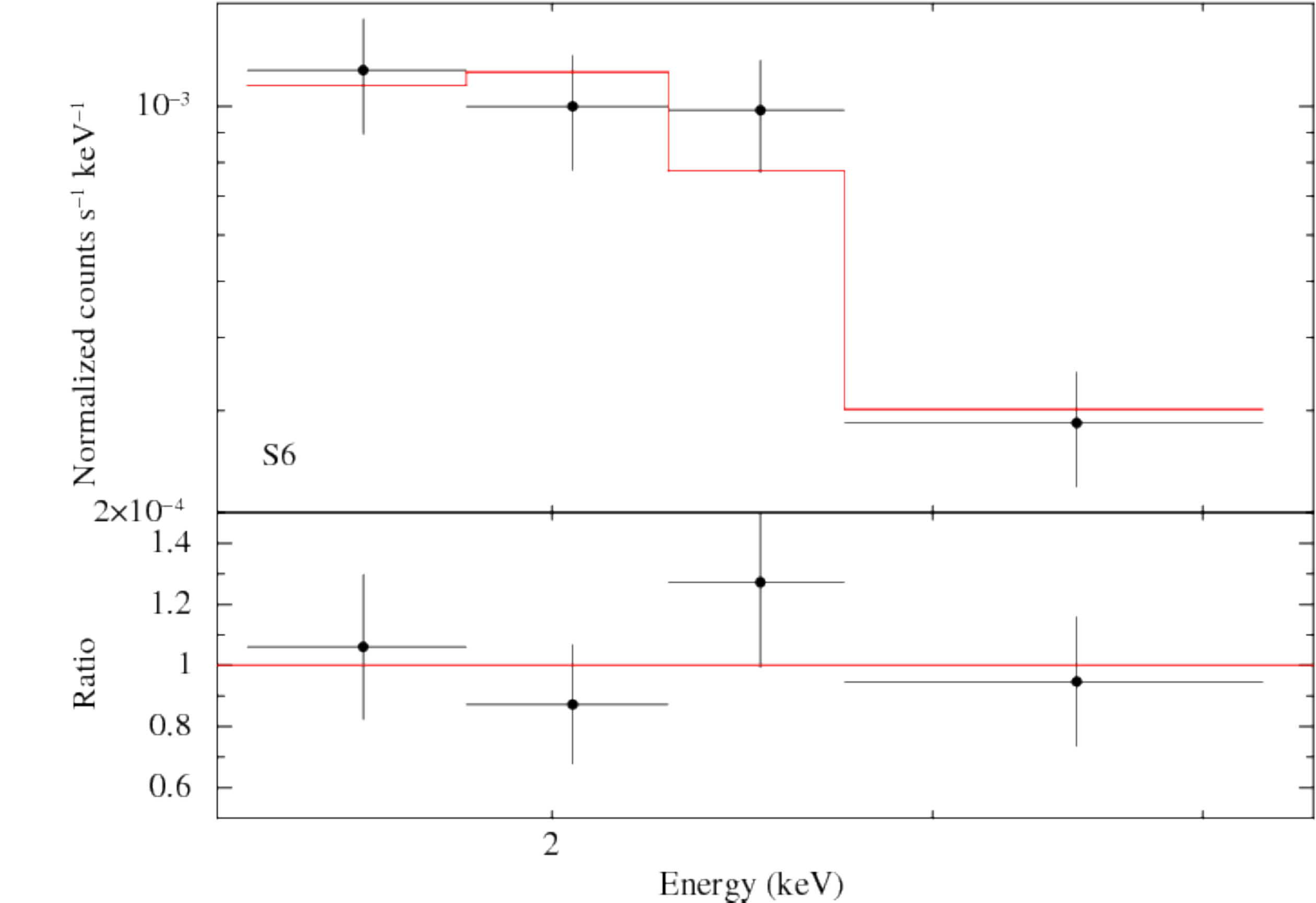}}
\caption{\label{spectrafig} Spectroscopy of the 4 brightest ULXs (S1, S4, S5, S6). For each source (as indicated in each plot), we show the spectrum, the best-fitting power-law model (solid line) and, in the bottom panel, the ratio to the model.}
\end{figure}

\begin{figure}
\centering
%\resizebox{0.95\hsize}{!}{\includegraphics[angle=-90]{sumfaint.ps}}
\resizebox{0.95\hsize}{!}{\includegraphics[angle=-90]{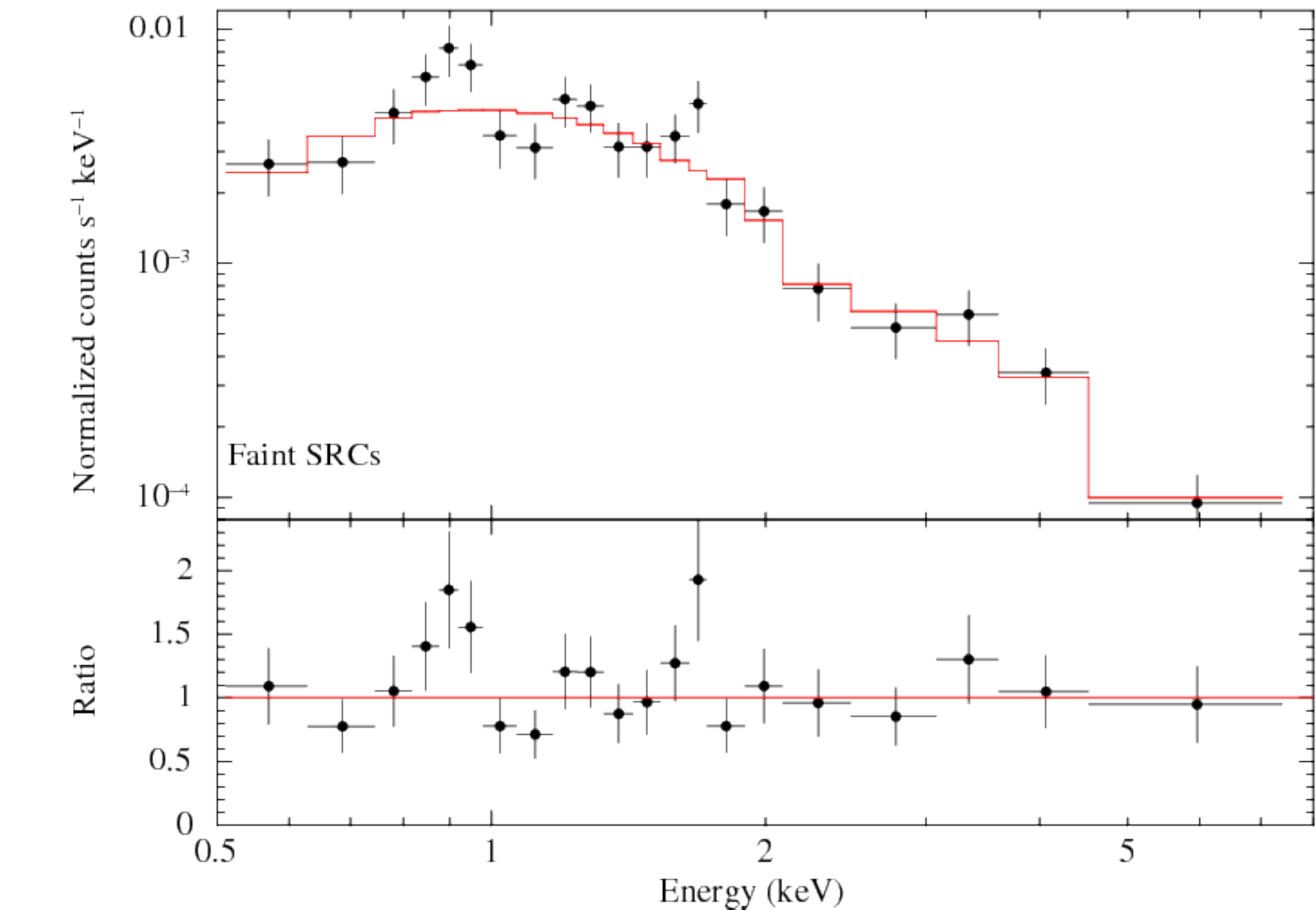}}
\caption{\label{spectrafigsum}Combined spectrum of the fainter sources (see Section\,\ref{anal}). The solid line indicates the best-
fitting power-law model and in the bottom panel we plotted the ratio to the model.}
\end{figure}

The few bright sources with enough counts ($\sim$100--400) to allow individual spectral fits are modeled by a simple power law modified for the interstellar absorption. A satisfactory description of their spectra has a hydrogen column density $N_{\mathrm{H}}$ of a few $10^{21}$~cm$^{-2}$ (the total Galactic \nh\ in the direction of NGC\,2276 is $\sim$$5.7\times10^{20}$~cm$^{-2}$; \citealt{kalberla05}) and slopes between $\Gamma = 1.4$ and 2.2. The observed normalised net count distribution with the best fit model is plotted in Fig.\,\ref{spectrafig}, while the fit results are listed in Table\,\ref{fit}. These spectra are consistent with typical results for ULXs when only low-count-statistics spectra are available (e.g. \citealt{swartz04}).
For the other sources, we considered the cumulative spectrum (see Section\,\ref{anal}) for a total of $\sim 300$ net counts. An absorbed power law can be fit also to these data, resulting in an absorbing column $N_{\rm H}=(1.7\pm0.9)\times10^{21}$~cm$^{-2}$ and photon index $\Gamma=1.9\pm0.4$ (see Fig.\,\ref{spectrafigsum}).
This mean spectrum was used to convert the observed count rates of the faintest sources into fluxes, see Table\,\ref{flux}.

\begin{table}
\centering
%\caption{Point source unabsorbed fluxes and luminosities computed from the power-law fit to the mean spectrum of the fainter sources (see Sections \ref{anal} and \ref{pntsrcs}). The four brightest sources have been also fitted individually (Table\,\ref{fit}) but we recalculated here their fluxes for comparison.
\caption{Point source unabsorbed fluxes and luminosities computed from the power-law fit to the mean spectrum of the fainter sources (see Sections \ref{anal} and \ref{pntsrcs}).
}
\label{flux}
\begin{tabular}{@{}lccc}
\hline                                    
Name &    $F_{\mathrm{X}}^{(0.5-10\,\mathrm{keV})}$  &    $L_{\mathrm{X}}^{(0.5-10\,\mathrm{keV})}$   &    $L_{\mathrm{X}}^{(2-10\,\mathrm{keV})}$    \\
     & ($10^{-14}$~\flux) & ($10^{39}$~\lum)  & ($10^{39}$~\lum) \\
%\hline
%S1   &  $8.68\pm0.43$    &    12.22   & 6.34\\   
%S4   &  $3.66\pm0.27$    &     4.73   & 2.67\\       
%S5   &  $7.04\pm0.38$    &     9.11   & 5.14\\    
%S6   &  $2.18\pm0.21$    &     2.81   & 1.59\\
\hline                                  
S2   &  $0.96\pm0.14$    &     1.24   & 0.70\\           
S3   &  $0.90\pm0.14$    &     1.16   & 0.66\\    
S8   &  $0.30\pm0.08$    &     0.38   & 0.22\\
S9   &  $0.17\pm0.06$    &     0.22   & 0.13\\                           
S10  &  $0.40\pm0.09$    &     0.52   & 0.29\\                         
S11  &  $1.42\pm0.17$    &     1.83   & 1.04\\                        
S12  &  $0.44\pm0.10$    &     0.57   & 0.32\\                    
S13  &  $1.11\pm0.15$    &     1.43   & 0.81\\                   
S14  &  $0.40\pm0.09$    &     0.52   & 0.29\\
S15  &  $0.61\pm0.11$    &     0.79   & 0.44\\
S16  &  $0.28\pm0.08$    &     0.36   & 0.20\\
\hline                                  
\end{tabular}                           
\end{table}                             
                                        
\begin{figure}
%\hbox{\includegraphics[width=8.3cm]{fig4.ps} }
\hbox{\includegraphics[width=8.3cm]{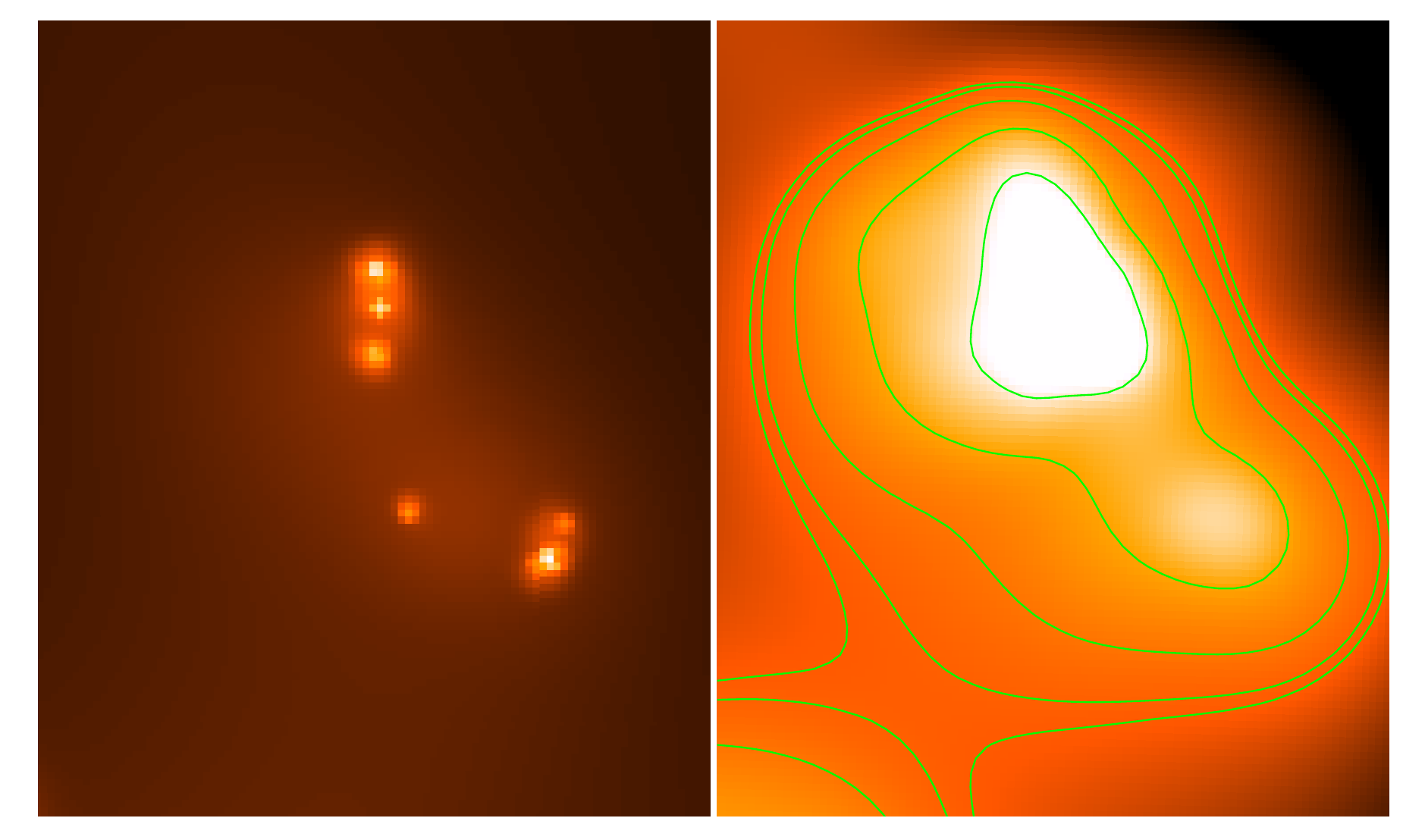} }
\caption{Field of the ULX candidate XMMU\,J072649.2+854555 as imaged by \cxo\ (left) and \xmm\ (right) in the 0.3-8 keV energy band. Each image side is 55 $\times $ 46 arcsec wide. The green contour levels are logarithmically spaced from 0.1 cts/sq.pix. Several distinct X-ray sources are resolved by \cxo.\label{ulx}}
\end{figure}

The number of sources with $L_{\mathrm{X}}^{(0.5-10\,\mathrm{keV})} \geq 10^{39}$~\lum\ is 8. \citet{liu11} classify 9 sources as ULXs, however the two lists have only the 5 brightest objects in common (S1, S4, S5, S6, S11). The faintest ones are not in the \citet{liu11} list since they use a cutoff luminosity of $2. \times 10^{39}$~\lum\, in the 0.3--8~keV band and a different distance of $D=36.8$~Mpc. The other sources in the \citet{liu11} list which are not listed here are: NGC2276-X4, which is the F8 star SAO 1148 (e.g. from the Simbad Astronomical Database\footnote{http://simbad.u-strasbg.fr/simbad/});  NGC2276-X6 which is outside the optical region of the galaxy ($D25 = 177\farcs1$ from NED) at $\sim 2^{\prime}$ from the nucleus of the galaxy; NGC2276-X7 which is on the border of the CCD and did not survive our thresholds; NGC2276-X9 which is on the next chip S2 (CCD6) which we did not even analyze. 
A comparison of \xmm\ and \cxo\ images of the field of the \xmm\ ULX candidate XMMU\,J072649.2+854555 is shown in Fig.\,\ref{ulx}. Instead of a single bright source, \cxo\ found 5 distinct sources (S1, S3, S4, S5, and S6; all in the ULX luminosity range, plus a possible other source near S1, which is visible also in Fig.\,\ref{images} but the detection algorithm is unable to pick it up) roughly following the \xmm\ isophotes (see also \citealt{wolter11}). The luminosity of these sources ranges from a few $10^{39}$ to a few $10^{40}$ \lum\ (Tables \ref{fit} and \ref{flux}). 

To investigate possible variability of the \cxo\ sources in the region of XMMU\,J072649.2+854555, we can compare the total luminosity from \citet{davis04} ($L_{\mathrm{X}}^{0.5-10\,\mathrm{keV}} = 1.1\times10^{41}d^2_{45.7}$~\lum$ =5.7\times10^{40}d^2_{32.9}$~\lum) to that of the region in the \cxo\ dataset corresponding to the \xmm\ point spread function (PSF), which we chose as a circle of 30$^{\prime\prime}$ radius centered at the \xmm\ position. This includes sources S1, S3, S4, S5 and S6, plus the unresolved sources and diffuse component. We extract a spectrum and fit it with the \citet{davis04} best fit model (\textsc{diskbb+mekal}), see their table 1) leaving the normalization free. We obtain an unabsorbed luminosity $L_{\mathrm{X}}^{0.5-10\,\mathrm{keV}} = 5.1\times10^{40}d^2_{32.9}$~\lum, which is consistent within uncertainties with the expectation.
We make the same comparison also for the other two sources detected in both the \cxo\ and the \xmm\ observation, namely source XMMU\,J072718.8+854636 (which corresponds to source S10), and source XMMU\,J072816.7+854436 (which corresponds to S15). If we consider the model of \citet{davis04} for each of the two sources, we find that XMMU\,J072718.8+85463 is apparently fainter by one dex ($L_{\mathrm{X}}^{0.5-10\,\mathrm{keV}} = 5.2\times10^{38}~\lum$) with respect to the expectation of $L_{\mathrm{X}}^{0.5-10\,\mathrm{keV}} = 4.2\times10^{39}d^2_{45.7}$~\lum$ =2.2\times10^{39}d^2_{32.9}$~\lum). However if we consider a region of 30$^{\prime\prime}$ around S10, we detect in \cxo\  158 $\pm$ 46 counts, from the diffuse contribution of the galaxy besides the 19.2 net counts from S10: therefore it is probable that the \xmm\ detection was similarly contaminated by the plasma component.
A different situation is found for XMMU\,J072816.7+854436 which is quite isolated: the \cxo\ luminosity for a power law spectrum is about a factor of 1.4 higher ($L_{\mathrm{X}}^{0.5-10\,\mathrm{keV}} = 7.9\times10^{38}~\lum$) than expected from \xmm\ : $L_{\mathrm{X}}^{0.5-10\,\mathrm{keV}} = 1.1\times10^{39}d^2_{45.7}$~\lum$ =5.7\times10^{38}d^2_{32.9}$~\lum, which however we deem not compelling given the uncertainties in spectral shape and the different response matrices of the two instruments. In conclusion, the possible small variation in flux between the observations, given the different spatial resolution of the instruments, cannot be confirmed statistically.

We note that source S6 is possibly coincident with a triple radio source of total $L_{5\,\mathrm{GHz}} = 9.51 \times 10^{36}$ erg s$^{-1}$, which \citet{mezcua13} suggest could be the first evidence of an extended jet in a non-nuclear source. 
The presence of jets in ULXs has been invoked in a few cases \citep[see also][]{cseh14} but the evidence is not firm yet. 
The source has a very blue (hard) color in Fig.\,\ref{truecolor} which might indicate an excess absorption. This however is not enough to imply that the source is a background source since high levels of absorption are known to be intrinsic to many ULXs.

\subsection{The X-ray luminosity function}
\begin{figure}
%\resizebox{\hsize}{!}{\includegraphics[]{xlf.eps}}
\resizebox{\hsize}{!}{\includegraphics[]{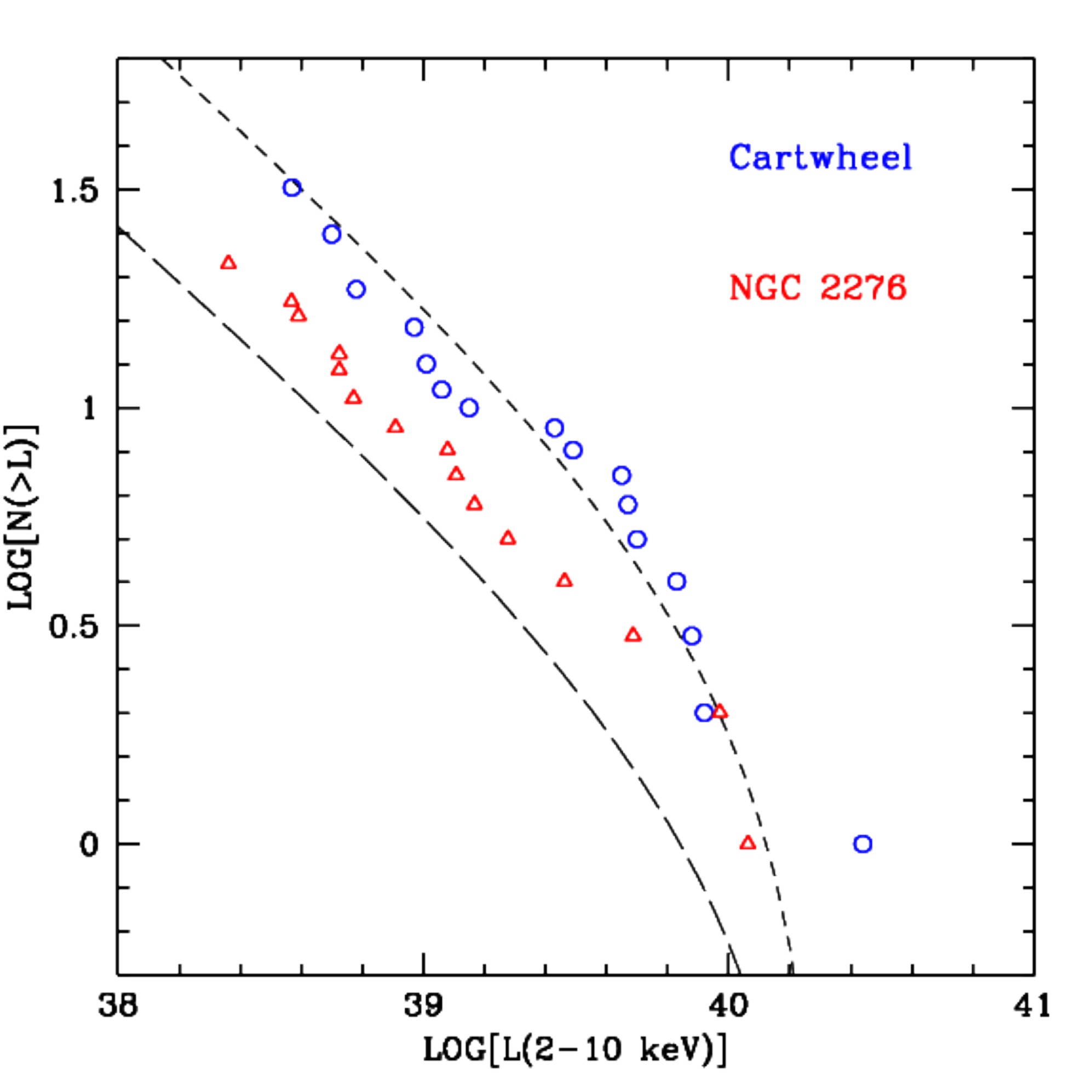}}
\caption{\label{xlf} The X-ray luminosity function of NGC\,2276 computed in the 2--10~keV band by using the sources detected in the galaxy area by \cxo. The nucleus has been removed, as well as one of the faint sources, to account for the expected contamination from background sources (see Section\,\ref{pntsrcs}). The solid lines represent the universal XLF for HMXB by \citet{grimm03}, normalized to 5 (long-dashed line) and 15~M$_{\sun}$~yr$^{-1}$ (short-dashed line).
}
\end{figure}

In order to construct the X-ray luminosity function (XLF) of NGC\,2276, we have to assess two possible sources of contamination: spurious, unwanted sources which add to the XLF, and missing sources due to detection uncertainties. We also do not include the central source (S7).
To account for the possible presence of spurious contamination from background sources among the \cxo\ point sources, we randomly excluded one of the three faintest sources, following the results of Section\,\ref{pntsrcs} for a chance coincidence of 1.4 source at the faintest fluxes. The contamination becomes less than 0.2 sources in the entire galaxy, as defined by the $D25$ diameter, at fluxes $F_{\mathrm{X}} \geq 10^{-14}$~\flux.
To correct for incompleteness, we base our assessment on fig. 12 from \citet{kf03} in which the detection probabilities as a function of background counts and source counts are plotted. The background includes both diffuse emission from the galaxy and field background and is estimated in the case of NGC\,2276 to be at a level of  0.15 counts pixel$^{-1}$ (by averaging total counts in the galaxy region, up to $D25$, after excluding detected sources). We use the second panel, relative to a 2$^{\prime}$ off-axis, since the whole galaxy is within this limit. We apply the interpolated correction to all sources with less than 22 counts. This is a conservative estimate. 
In Fig.\,\ref{xlf}, we compare the resulting XLF with that of the Cartwheel galaxy, one of the richest galaxies in ULXs \citep{wolter04}. The Cartwheel XLF was corrected with the same procedure: in this case the background is lower (0.05 counts pixel$^{-1}$) and therefore we corrected only sources with less than 10 counts.
Just to give an indication, the slope of the resulting XLF is 
 $\alpha_{\mathrm{N}} = -0.71\pm0.03$ for NGC\,2276 and $\alpha_{\mathrm{C}} = -0.75\pm0.05$ for the Cartwheel. These numbers become lower ($\alpha_{\mathrm{N}} = -0.66$ and  $\alpha_{\mathrm{C}} = -0.60$) if we restrict the fit to the $\log L_{\mathrm{X}} \leq 39.8$, consistent with the `universal' luminosity function slope of  $\alpha = -0.60$ found by e.g. \citealt{grimm03,swartz11,mineo12a} for high mass X-ray binaries (HMXBs) and proposed to scale with the SFR.
The XLF from \citet{grimm03} is also plotted in Fig.\,\ref{xlf} for two different values of the SFR of 5 and 15~M$_{\sun}$~yr$^{-1}$. Also, by comparing the total X-ray luminosity or the number of ULXs with the formulas of \citet{grimm03}, similar results are obtained: SFR = 5 and 5.5~M$_{\sun}$~yr$^{-1}$, respectively. These values are in the same range of SFR derived in other wavebands (see Sect.\,\ref{Intro}). Also eq. (20) from \citet{mineo12a} gives a predicted number of sources with $L_{\mathrm{X}} > 10^{38}$~\lum$ =3.2\times\mathrm{SFR} \geq 16$, consistent with the number of sources found, taking into account that at least one is spurious and the incompleteness of detection at the limit. 
%%%%%{\bf The results on the number of ULXs per SFR should also be compared to the results of the \citet*{smith12} survey of ULXs in Arp Atlas galaxies. }

\section{The diffuse emission component}\label{sec:diffuse}

\begin{figure}
%\resizebox{\hsize}{!}{\includegraphics[]{prof2+psf.eps}}
\resizebox{\hsize}{!}{\includegraphics[]{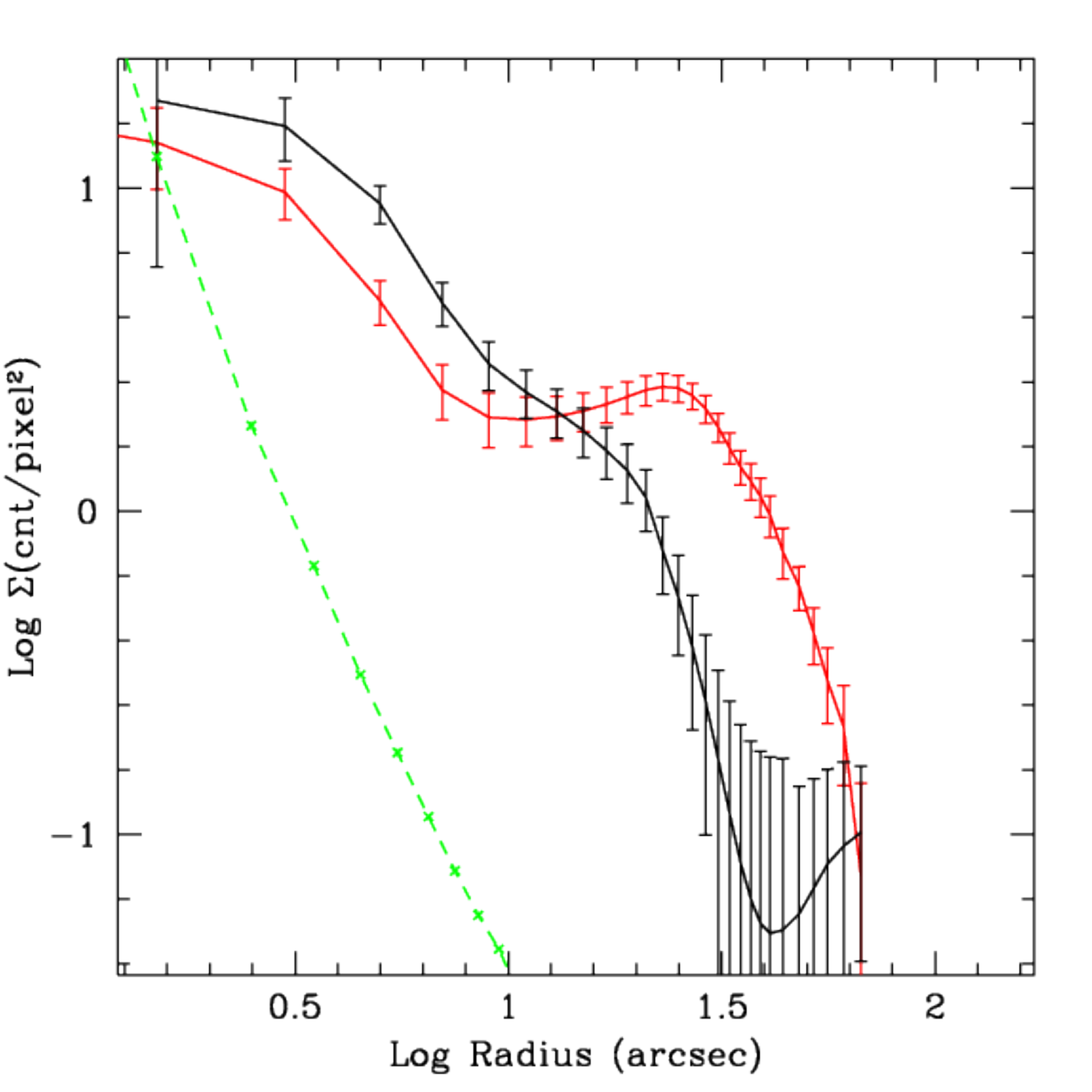}}
\caption{Profile in the direction 280--370$\degr$, in red vs. the opposite one: 110--200$\degr$, in black. The green dashed line indicates the extent of the PSF computed via \textsc{chart}, normalized to the first data point.\label{profile}}
\end{figure}

The nuclear source of NGC\,2276 was previously detected with \rst\  as variable \citep{davis97} at $L_{\mathrm{X}}^{(0.5-2\,\mathrm{keV})} \sim 2\times10^{40}d^2_{45.7}=10^{40}d^2_{32.9}$~\lum. Notably, the source was not detected in the 2004 \xmm\ observation, which set an upper limit of  $L_{\mathrm{X}}^{(0.5-10\,\mathrm{keV})} < 1.2\times10^{39}d^2_{32.9}$~\lum\ on its emission.
However, a correct comparison of  the flux would need a detailed modeling of the surrounding emission, given the different resolution of the instruments that have observed NGC\,2276 through the years.  
Source S7, coincident with the optical nucleus of the galaxy, is extended as we show in Fig.\,\ref{profile}, where we compare it with the \cxo\ PSF derived with \textsc{chart} with standard procedures at the position of the nucleus.
We plot two different profiles in the NE and SW directions (110 to 200$\degr$ and 280 to 370$\degr$, respectively) to show the different extent at large radii in the two directions, as already noted by \citet{rasmussen06}.

Since the diffuse emission of NGC\,2276 and in its proximity was studied in detail by \citet{rasmussen06}, we just verify the consistence of our results for the diffuse emission of the disc (after removing the point sources) for their region A (see their figure~2). We used an extraction region as similar as possible to that of \citet{rasmussen06}: a circle of 50$^{\prime\prime}$ radius, visible in Fig.\,\ref{images}. 

\begin{figure}
\centering
%\resizebox{0.95\hsize}{!}{\includegraphics[angle=-90]{gas_spectrum.ps}}
\resizebox{0.95\hsize}{!}{\includegraphics[angle=-90]{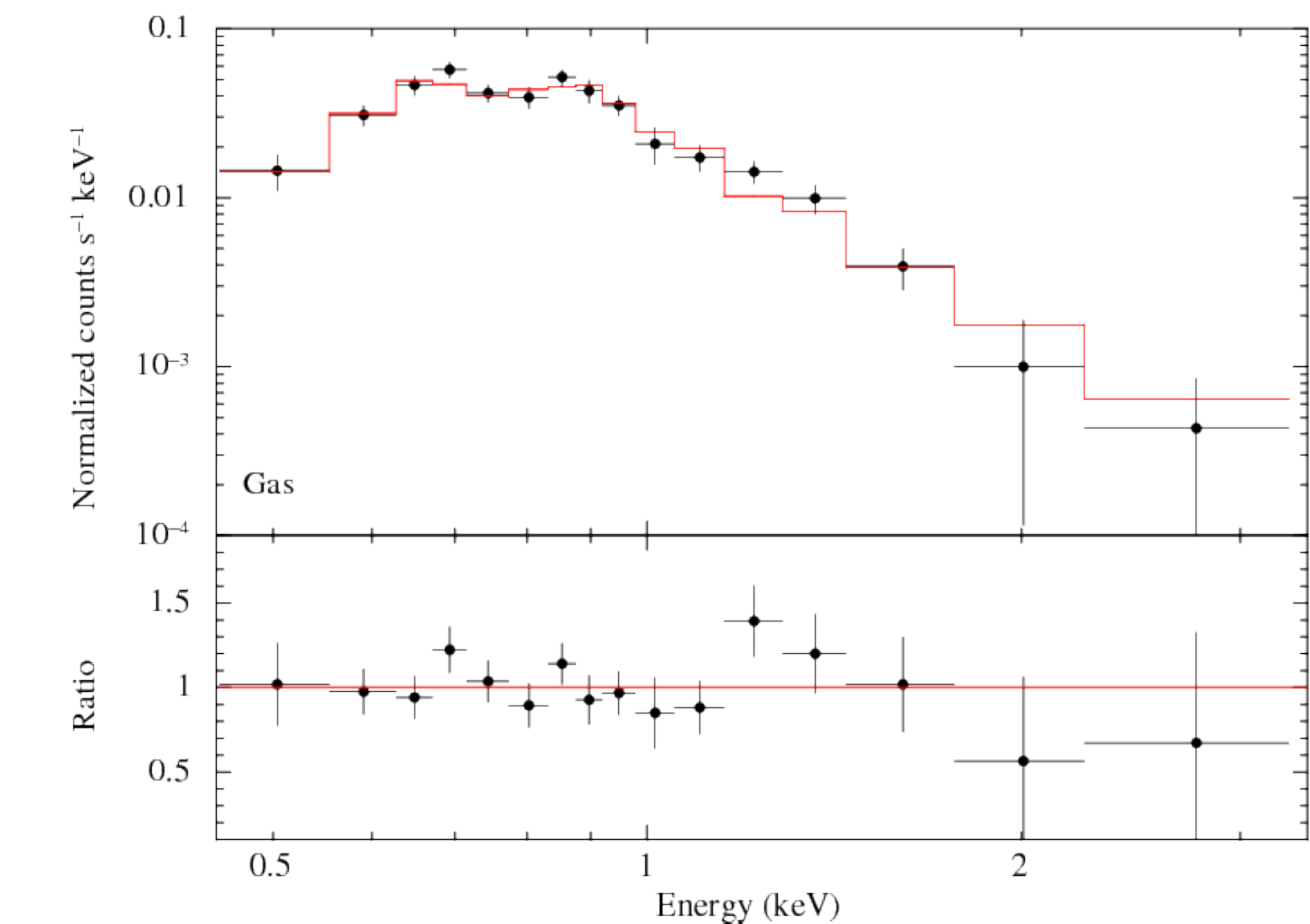}}
\caption{\label{spectrafiggas}Spectrum of the central region of the galaxy (see Section\,\ref{sec:diffuse}). The solid line indicates the best-fitting power-law model and in the bottom panel we plotted the ratio to the model.}
\end{figure}

We make a more general hypothesis that there might be intrinsic absorption in the galaxy, and we bin the spectrum so as to have a minimum of  100 counts per bin, but we checked that the values are not very dependent on binning. In this way, in addition to the Galactic absorption, we measured an absorbing column equivalent to $N_{\rm H} = (4.2^{+1.7}_{-1.3}) \times 10^{21}$~cm$^{-2}$ for the best fit ($\chi^{2}_\nu= 1.01$ for 13 degrees of freedom (dof)). The resulting spectrum is plotted in Fig.\,\ref{spectrafiggas}.
This results in a lower temperature of $kT=0.18^{+0.05}_{-0.02}$~keV and an only poorly constrained photon index $\Gamma =4\pm2$, but we note that the power-law component is necessary to properly fit the spectrum and we freeze it to $\Gamma = 1.7$, appropriate to represent both a low luminosity active galactic nucleus (AGN) component and/or the unresolved point source component. The corresponding total luminosity is $L_{\mathrm{X}}^{(0.5-2\,\mathrm{keV})} = 1.8\times 10^{41}$~\lum. The luminosity of the power-law component is $L_{\mathrm{X}}^{(0.5-2\,\mathrm{keV})} = 2.1 \times 10^{40}$~\lum, about 12 per cent of the total luminosity and consistent with the expectation for the unresolved point source component derived from the $K$ band luminosity of NGC\,2276  $L_K = 10.7$, which corresponds to $L_{\mathrm{X}}^{2-10\,\mathrm{keV}} \sim 10^{40}$ erg s$^{-1}$, by using the relation in \citet{kf04}. Since there should also be a fraction of unresolved HMXBs, this leaves little room for a powerful AGN.

We observe that our revised temperature and luminosity have essentially no impact on the results of \citet{rasmussen06} (actually, the higher luminosity and lower temperature enforce their conclusion that diffuse X-ray emission from the western edge does not arise from a shock-compressed gas component).

If we compute the total emission from the $D25$ of the galaxy, then the diffuse component is
$L_{\mathrm{X}}^{0.5-2\,\mathrm{keV}} \sim 2.5 \times 10^{41}$ erg s$^{-1}$: we can compare it with the findings of \citet{mineo12b} that $L_{\mathrm{X}}^{\mathrm{diffuse}}/SFR = 1.5 \times 10^{40}$ erg s$^{-1}$ / (M$_{\sun}$~yr$^{-1}$), albeit with a large scatter of $\sim$ a dex and very model dependent. Again this gives a SFR in the range 5--$25$ M$_{\sun}$~yr$^{-1}$.

%% {\bf why don't they give an estimate of the Lx(diffuse)/SFR from the whole galaxy, and compare it with previous works, such as Smith, Soria, Struck, et al. (2014, AJ, 147, 60) and Mineo et al. (2012, MNRAS, 426,1870)?}
%% NB: counts in 50'' are 

We recall also that the source is immersed in an IGM component: we report the parameters found by \citet{davis96} using \rst\ data, rescaled to our distance, which we will exploit for the subsequent simulations. They found an extent of about 25$^{\prime}$ (0.24~Mpc) for the X-ray emitting IGM, centered close to, but not exactly on, the central elliptical galaxy NGC\,2300. The surface brightness of the medium is described by a King function with core radius $r_0 \sim 4.3^{\prime}$ and index $\beta  = 0.41$, and has a luminosity of $L_{\mathrm{X}}^{\mathrm{bol}} = 1.12 \times 10^{42}d^2_{45.7}=5.8\times10^{41}d^2_{32.9}$~\lum. The mass of the hot intragroup gas is $M_{\mathrm{gas}} = (1.25 \pm 0.13) \times 10^{12}$~M$_{\sun}$. 
\citet{rasmussen06} find similar values, but the \rst\ larger field of view is better suited to measure this large extent component.

\section{Where does the morphology of NGC\,2276 come from?}\label{s-simul}
%self\subsection{An order-of-magnitude estimate}
%{\bf Michela:  in particular an old version of SPH, gadget2. Indeed it is known that the SPH gadget2 has difficulties in capturing the correct  gas dynamics in group or cluster environment. I would like the authors to comment on this crucial point in the paper. }
Whether the perturbed morphology of NGC\,2276 is the result of tidal interactions with NGC\,2300, or of ram pressure stripping is still an open question. The simplest way to estimate the importance of tidal forces, without running a simulation, is to compare the centripetal acceleration of NGC2276 ($a_{\mathrm{c}}$) to the radial acceleration ($a_{\mathrm{r}}$) exerted by the other galaxies in the group, which is dominated by NGC2300 and by the hot gas component:
\begin{eqnarray}\label{eq:tid}
a_{\mathrm{c}}(r)=\frac{G\,M(r)}{r^2},\\
a_{\mathrm{r}}(D) = G\,M_{\mathrm{group}}(D)\,\left[\frac{1}{D^2}-\frac{1}{(D+r)^2}\right]\label{eq:tid2},
\end{eqnarray}
where  $r$ and $M(r)$ are the three-dimensional distance from the center of NGC\,2276 and the mass of NGC\,2276 inside $r$, respectively, while $M_{\mathrm{group}}(D)$ is the mass of the group enclosed within a distance $D$ from the center of NGC\,2300.

To derive the actual distance between NGC\,2276 and NGC\,2300 we used the velocity of NGC\,2276 with respect to the IGM, derived assuming that the IGM is coupled to NGC\,2300.  \citet{rasmussen06} derive a Mach number of $\mathcal{M} = 1.70\pm 0.23$, corresponding to a three-dimensional velocity of $v= v_\mathrm{s} \times \mathcal{M} = 865 \pm 120$~km s$^{-1}$. The distance $D = x/cos \alpha$ is then derived by measuring $x =$ 83.2 kpc (the distance between the centers of the two galaxies projected on the plane of the sky) and $\cos\alpha = \Delta v_{\mathrm{r}} /v$, where  $\Delta v_\mathrm{r} = 421$~km s$^{-1}$ is the relative recession velocity between the two galaxies \citep{rasmussen06}. The three-dimensional distance between the two galaxies is then $170$~kpc, under these assumptions.

 For $r=13$ kpc (which is approximately the isophotal radius of NGC\,2276) and $M(13\,{\rm kpc})=2\times10^{11}$ M$_{\sun}$, we obtain $a_c=1.6\times10^{-8}$ cm s$^{-2}$. For $D=170$ kpc,
%(which is the current three-dimensional distance between NGC\,2276 and NGC\,2300), 
$a_r(D)\sim5\times10^{-10}$ cm s$^{-2}$. 
Thus, we expect tidal forces to be small today, when compared with the self-gravity of the galaxy. 
Let us consider now the importance of ram-pressure stripping. The distance from the center of a galaxy at which ram-pressure stripping becomes important can be evaluated as (\citealt{yoshida08})

\begin{equation}
r_{\mathrm{rp}}=\frac{h}{2}\ln\left({\frac{G\,M_\ast\,M_{\rm ISM}}{2\,\pi v^2\rho_{\rm IGM}\,h^4}}\right),
\end{equation}

where $h$ is the scale-length of the disc, $M_{\ast}$ and $M_{\mathrm{ISM}}$ are the total stellar mass and inter-stellar medium mass of the galaxy, $v$ is the relative velocity between the galaxy and the IGM, and $\rho_{\mathrm{IGM}}$ is the local density of IGM. If we assume $h=3$ kpc,   $M_{\ast}=2.8\times10^{10}$~M$_{\sun}$, $M_{\mathrm{ISM}}=8\times10^9$~M$_{\sun}$, $v=850$~km s$^{-1}$ and $\rho_{\mathrm{IGM}}=5\times10^{-27}$~g cm$^{-3}$, which are consistent with the observations of NGC\,2276 (\citealt{davis96,rasmussen06}) and its environment (we use the same values for the simulations described in the next Section), we obtain  $r_{\mathrm{rp}}=6.5$~kpc, which is larger than $h$ but still inside the isophotal radius of  NGC\,2276 (13.2 kpc). Thus, we expect ram-pressure to be quite effective in perturbing the morphology of NGC\,2276. 

Even if from equations (\ref{eq:tid}) and (\ref{eq:tid2}) we have shown that tidal forces are not as efficient as ram pressure, it is reasonable to expect that there is an interplay between tidal forces and ram pressure, leading to an enhancement of  NGC\,2276 stripping with respect to the case in which ram pressure was the only process at work (see e.g. the argument in \citealt{yoshida08}). In addition, viscous stripping (e.g. \citealt{nulsen82}) might also play a role, removing matter from the ISM.

\section{Numerical simulations}
To study the morphology of NGC\,2276 in more detail than with our order-of-magnitude calculations, we ran N-body/smoothed-particle hydrodynamics (SPH) numerical simulations of the interaction between NGC\,2276 and its environment. We used the \textsc{gadget-2} code \citep{springel01, springel05}. 

The initial conditions are taken to the simplest possible level. In all the simulations, we put NGC\,2276 on a parabolic orbit around NGC\,2300 (assumed to sit at the bottom of the gravitational potential well of the group). The two galaxies start from an initial separation of $170$~kpc, $85$~Myr before the periapsis. At periapsis, the galaxies reach their minimum distance of $160$~kpc, and after another $85$~Myr they reach their present separation of $170$~kpc. We ran the simulations for 1 Gyr, but the most relevant effects occur during (and within $\lesssim200$ Myr after) the periapsis passage. 

We describe the elliptical galaxy NGC\,2300 as a rigid potential, to reduce the computing time. We further assume that the total contribution is due to the dominant component of dark matter, described by a Navarro, Frenk and White (NFW) profile \citep{navarro97}:
\begin{equation}
\rho(x)=\frac{\rho_0}{x\,(1+x)^2},
\end{equation}
where $x=\frac{r}{r_{\mathrm{s}}}$, with $r_{\mathrm{s}}=R_{200}/c$ and $R_{200}=(M_{\mathrm{tot}}\,G/H_0^2)^{1/3}$  ($H_0=73$ km s$^{-1}$ Mpc$^{-1}$ being the Hubble constant, $M_{tot} $ being the total mass and $c$ being the concentration parameter), and
 \begin{equation}
   \rho_0=\frac{200\,\rho_{\rm crit}\,c^3}{3\,\left(\ln{(1+c)}-c/(1+c)\right)},
\end{equation}
where $\rho_{\rm crit}$ is the current critical density of the Universe. We assume $M_{\mathrm{tot}} = 1.3 \times 10^{13}$~M$_{\sun}$ and concentration parameter $c=10$. 

The second ingredient is the IGM. We perform three different simulations for different choices of the IGM profiles. The first is a `dry' simulation, with no IGM. This is a control simulation aiming to isolate the purely tidal effects between the galaxies. 

The second and third simulations assume that the IGM is present, but with  different distributions. In both cases the distributions are centered on NGC\,2300.  In both simulations, the gas is assumed to be adiabatic, with initial temperature $2\times10^6$ K, the total mass of gas is $M_{\mathrm{g}} = 1.25 \times 10^{12}$~M$_{\sun}$ and the mass of a single gas particle is $10^6$~M$_{\sun}$ (for a total number of $1.25\times 10^{6}$ particles).

In the second simulation, the IGM has a NFW profile\footnote{The NFW profile usually refers to dark matter, and not to baryonic matter. The rationale of adopting a NFW profile for the baryonic IGM is the following. For large radii, the NFW has the asymptotic form $\varrho\propto r^{-3}$ to be  compared with the $\varrho\propto r^{-2}$ of an isothermal sphere. The NFW would model an IGM distribution concentrated close to NGC\,2300, while the isothermal sphere models a more widespread IGM.} with concentration $c=10$.

Finally, in the third simulation the gas is distributed as a King model \citep{king62}:
\begin{equation}
\rho(r)=\rho_0\,\left[1+\left(\frac{r}{r_{\mathrm{c}}}\right)^2\right]^{-3\,\beta+0.5},
\end{equation}
where $\rho_0=3.1\times10^{-3}$ cm$^{-3}$ (from \citealt{davis96}), $\beta=0.41$ (from \citealt{rasmussen06}) and $r_{\mathrm{c}}=4.0^{\prime}$ (from \citealt{rasmussen06}).

The third ingredient is a model of NGC\,2276. We parametrize it, according to \citet{wd05}, as the sum of three different components: bulge, disc and dark matter halo. The total number of particles in the dark matter halo, bulge and disc of the NGC\,2276 model are $2\times10^6$, $10^5$ and $6\times10^5$, respectively. The softening length is 100 pc. 

The dark matter halo is modeled as a NFW potential with $M_{\mathrm{tot}} = 6 \times 10^{11}$~M$_{\sun}$ and $c=10$. The bulge is represented by a S\'ersic profile, according to the \citet{prugniel97} deprojection $\rho_{\mathrm{b}}(r)\propto(r/r_{\mathrm{e}})^{-p}\,\exp{(r/r_{\mathrm{e}})^{1/n}}$, where we assume $r_{\mathrm{e}}=0.83$ pc, $p=-1$ and $n = 1.83$. The normalization of the profile is set by the total mass of the bulge $M_{\mathrm{b}}=6\times10^9$~M$_{\sun}$.

 The gas and the stellar component are distributed as an exponential Hernquist disc \citep{hernquist93}:
\begin{equation}
\rho_{\mathrm{d}}(R,z)=\frac{M_{\mathrm{d}}}{4\pi h^2z_0}\,\exp\left(s{-R/h}\right)\,\textrm{sech}^2\left({z/z_0}\right),
\end{equation}
where $R$ and $z$ are the cylindrical coordinates, $M_{\mathrm{d}} = 3.6 \times 10^{10}$~M$_{\sun}$ is the total mass of the disc, $h=3$ kpc is the scale-length of the disc and $z_0=0.3$ pc is the scale-height of the disc. The disc is composed of both gas and stellar particles, with the total mass of gas and stars being $M_{\mathrm{d,\,g}}=0.8\times10^{10}$ and $M_{\mathrm{d},\,\ast}=2.2\times10^{10}$~M$_{\sun}$, respectively.

The gas component of the NGC\,2276 model is assumed to be adiabatic, with initial temperature of $T = 2 \times 10^4$~K \citep{rasmussen06}. We do not introduce a cooling function, but verify `a posteriori' if the cooling time $t_{\mathrm{cool}}$ is shorter than the dynamical free fall time $t_{\mathrm{dyn}}$ to estimate the rate of star formation. In particular, we estimate the cooling time as
\begin{equation}
t_{\mathrm{cool}}=\frac{3}{2}\,\frac{n\,k\,T}{n_e\,n_p\,\Lambda(Z,T,n)},
\end{equation}
where $n$ is the total gas density, $n_e$ is the density of electrons, $n_p$ is the density of protons, $k$ is the Boltzmann constant, $T$ is the temperature, and $\Lambda(Z,T,n)$ is the cooling function, as a function of $n$, $T$ and of the metallicity $Z$. We used the cooling function of \citet{sutherland93}.

We assume that gas is able to form stars if 

\begin{equation}\label{eq:cool}
t_{\mathrm{cool}}<t_{\mathrm{dyn}}<t_{\mathrm{life}}, 
\end{equation}
where $t_{\mathrm{life}}$ is the life time of a gas clump or structure, derived directly from the simulations. 
%%%% NOSFR %%%% We assume that gas particles that satisfy this condition and that have a density $n>0.1$ atoms cm$^{-3}$ can form stars, with an efficiency that enforces the Schmidt law.%{\bf AW: Michela, non capisco ``enforces''} This method gives us a rough %an 'order-of-magnitude' 
%%%% NOSFR %%%% estimate of the SFR, even if the feedback from actual gas cooling and star formation is not accounted for. 
%{\bf MM: ma in realt\`a mattia non ha fatto nulla di tutto questo, se quello che ha scritto coincide con quello che ha fatto. leggendo l'ultima versione della tesi penso che il metodo sia proprio sbagliato concettualmente..}

Here below we shortly summarize the results of our simulations.

\subsection{Dry simulation}
Our order-of-magnitude estimate shows that tidal forces should not be very important for NGC\,2276. On the other hand, even in the simulation without IGM we observe the formation of (not strongly pronounced) tidal arms in both the stellar and the gaseous components.
In addition, the gas disc thickens. The regions where condition~\ref{eq:cool} is satisfied are very rare in the dry simulation, 
%%%% NOSFR %%%% and the predicted SFR does not exceed $1$ M$_{\sun}$ yr$^{-1}$, even during the periapsis passage
and we expect that star formation is not enhanced by the interaction, even during the periapsis passage.

No gas is stripped from the galactic disc, and the tails observed by \citet{rasmussen06} are not reproduced.

\subsection{NFW simulation}
\begin{figure}
\centering
%\resizebox{0.82\hsize}{!}{\includegraphics{NFW_300_zall_star.ps}}
\resizebox{0.82\hsize}{!}{\includegraphics{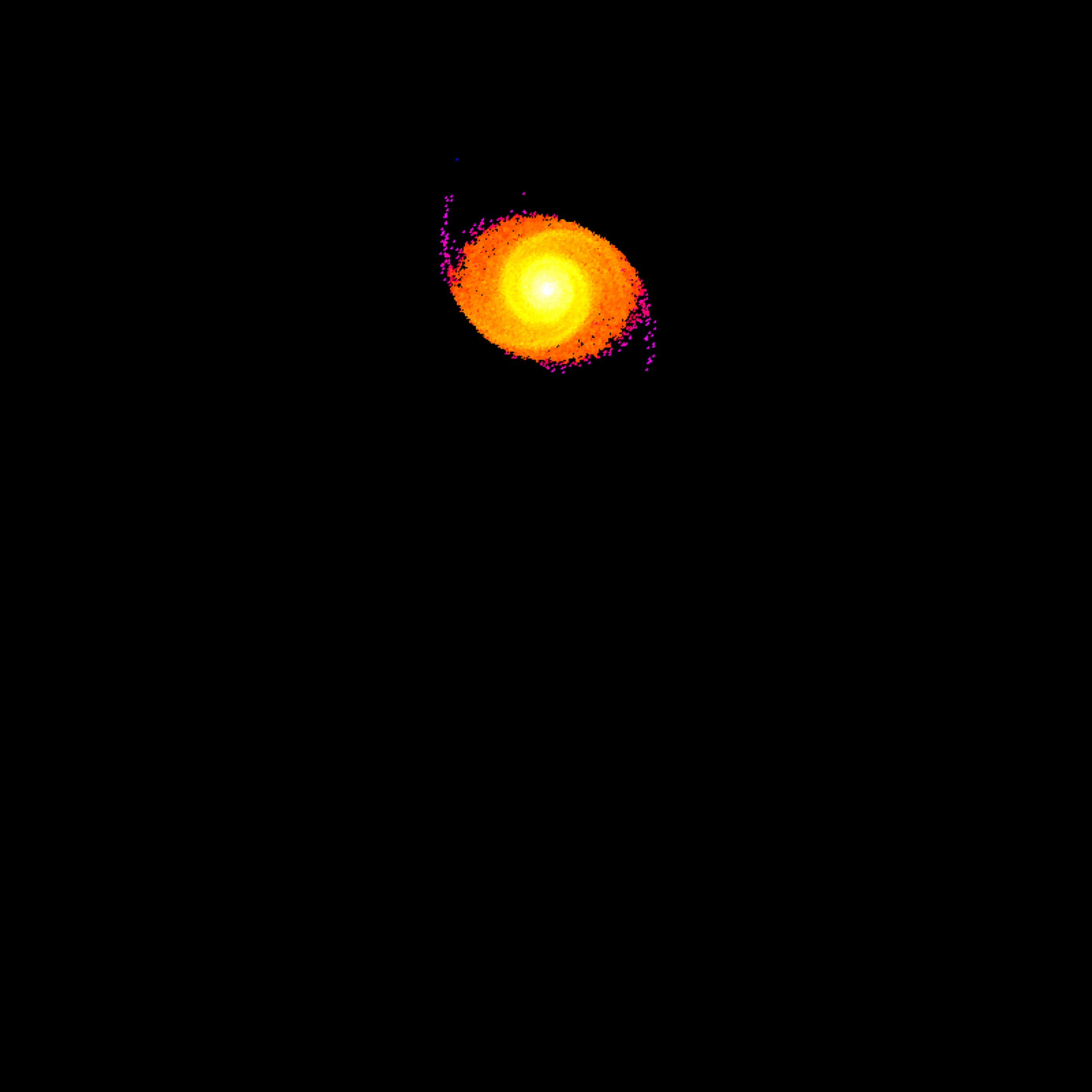}}
\vskip 10pt
%\resizebox{0.82\hsize}{!}{\includegraphics{NFW_300_zall_gas.ps}}
\resizebox{0.82\hsize}{!}{\includegraphics{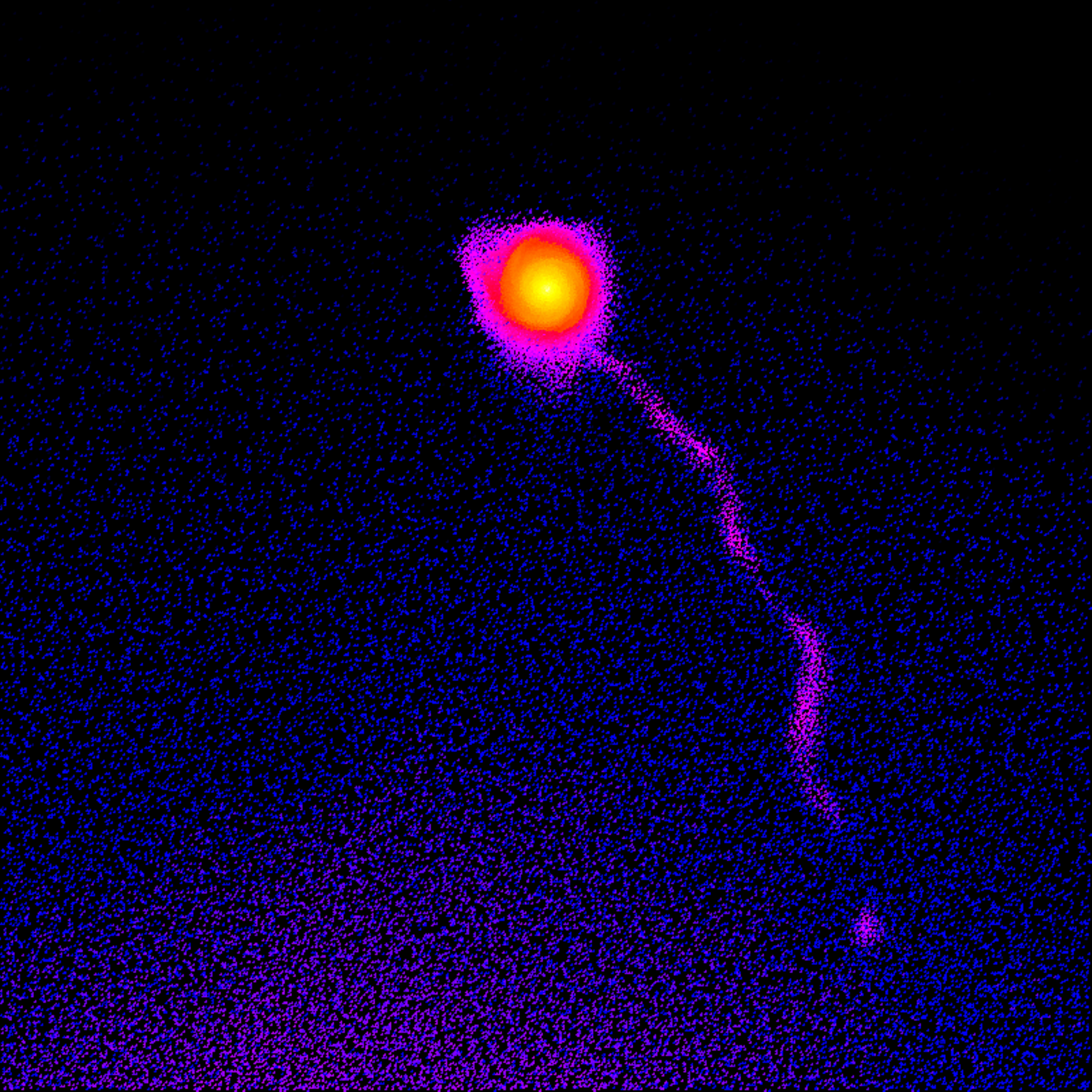}}
\vskip 10pt
%\resizebox{0.82\hsize}{!}{\includegraphics{NFW_300_yall_gas.ps}}
\resizebox{0.82\hsize}{!}{\includegraphics{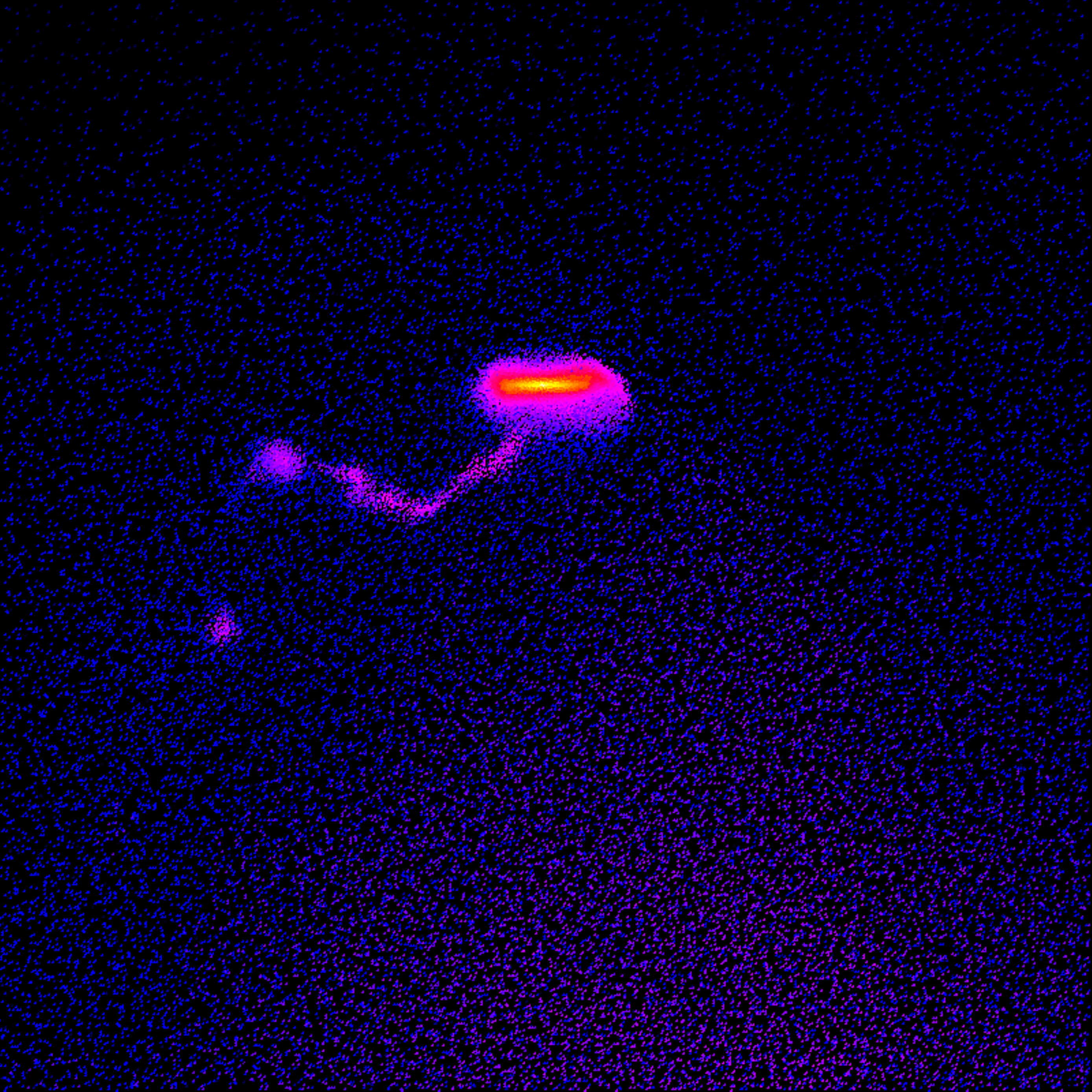}}
\caption{\label{fig:simula} Color-coded density map stars ({\it top panel}) and gas ({\it central and bottom panel}) at  $t=300$~Myr in the NFW simulation. The boxes measure 300 kpc per edge. Top panel: the density of stars has been projected along the z-axis; {\it central} ({\it bottom}) panel: the density of gas has been projected along the z-axis (y-axis). A tail structure is well visible in the {\it central} and {\it bottom} panel.
}
\end{figure}

The evolution of the disc galaxy changes significantly, if the IGM is accounted for. As we expect from our order-of-magnitude estimate, ram-pressure stripping and viscous stripping are efficient, and lead to the formation of tails already during the periapsis passage (see Fig.\,\ref{fig:simula}).
We also see that the front side of the galaxy disc is perturbed, recalling the asymmetric `bow-shock-like' shape of NGC\,2276.

The total mass in the tails is $\sim$$6\times10^8$ M$_{\sun}$ after the first periapsis passage (it will rise to $\sim$$1\times10^9$ M$_{\sun}$ at the end of the simulation). A part of this mass forms clumps that are visible in Fig.\,\ref{fig:simula}. The cooling time in the galaxy disc is $\sim$$10^{4-5}$ yr, while it is 10--20 Myr in the ram-pressure stripped tails. The dynamical time $t_{\mathrm{dyn}}$ is 200 Myr and 10 Myr in the galaxy disc and in the tails, respectively. Thus, star formation occurs in the galaxy disc and might occur also in the tails. 
%%%% NOSFR %%% We expect a SFR$\sim2$ M$_{\sun}$ yr$^{-1}$ (a factor of two lower than observed) in the first 150 Myr of the simulations, then the star formation fades away.

Finally, a small but significant fraction of the gas component of the NGC\,2276 model ($8.6\times10^8$ M$_{\sun}$, i.e. 6.6 per cent of the entire gas component) is completely stripped and lost from the galaxy during the entire simulation. 
%%%% NOSFR %%%% Fig.~\ref{massloss}  shows that most of this mass is lost during the first 100 Myr.

\subsection{Isothermal sphere simulation}
The third simulation is very similar to the second one. Even in this simulation, ram pressure produces a deformation in the front side of the galaxy disc and leads to the formation of tails and clumps. The total mass in the tails is $6\times10^8$ M$_{\sun}$ after the periapsis passage, and raises to $7.5\times10^8$ M$_{\sun}$ at the end of the simulation, with no significant difference with respect to the second simulation. In the tails, $t_{\rm cool}\sim t_{\rm dyn}\sim10$ Myr, suggesting that star formation is possible. In the disc galaxy, $t_{\rm cool}\sim10^{4-5}~\mathrm{yr} << t_{\rm dyn}\sim300$ Myr, indicating that star formation should occur.
%%%% NOSFR %%%% We estimate a rate of $\approx1$ M$_{\sun}$ yr$^{-1}$.
%\begin{figure}
%\centering
%\resizebox{\hsize}{!}{\includegraphics{massloss.eps}}
%\caption{\label{massloss} Gas mass loss in M$_{{\sun}}$ as a function of time in Myr from out model of NGC\,2276.}
%\end{figure}

The total mass of gas that is lost from the disc galaxy by the end of the simulation is $1.5\times10^8$ M$_{\sun}$, i.e. $\sim12$ per cent of the entire initial gas mass in the NGC\,2276 model. This is nearly a factor of two larger than in the second simulation. The difference can be explained with the fact that the IGM density goes as $\rho_{\rm IGM}\propto r^{-3}$ and $\propto r^{-2}$ in the second and in the third simulation, respectively. Since the density of the IGM at large distances from the elliptical galaxy in the third simulation is higher than that in the second simulation, we expect that ram pressure stripping will last longer than in the second simulation (i.e. it is important even when the disc galaxy is far from periapsis).

\subsection{Summary of simulation results and caveats}
In conclusion, we found that ram-pressure stripping and viscous stripping are the dominant processes and can lead to both the formation of the tails and the deformation of the front side of the disc galaxy. Tidal forces contribute marginally, by producing tidal arms and thickening the gaseous disc. We found a factor of 2--5 lower SFR than in the observations, but we are aware that our very approximate treatment of gas underestimates both cooling and SFR.

Moreover, the version of the SPH technique adopted in \textsc{gadget-2} is known to poorly resolve dynamical instabilities and mixing (e.g. \citealt{agertz2007}). The main reason is that the SPH formulation adopted in \textsc{gadget-2} inaccurately handles the case of low-density regions in contact with high-density ones. As a consequence, gas turbulence is artificially suppressed, and gas stripping due to ram pressure might be underestimated with respect to other formalisms (e.g. grid codes). In particular, we expect that the gas mass in the stripped tails is underestimated in our simulations, and the morphology of the resulting tails is quite smoother than it should be. This implies that our results can be regarded as lower limits for the ram-pressure effects. 
Bearing these {\it caveats} in mind, we conclude that our simulations are able to qualitatively reproduce the peculiar shape of NGC~2276, and to confirm the importance of ram pressure and viscous stripping.

\section{Summary \& Conclusions}

We have analyzed a \cxo\ observation of NGC\,2276, a spiral galaxy in the group of the elliptical NGC\,2300. We have paid particular attention to the point sources detected in the area of the galaxy. We find that a large number of ULXs are present, which makes NGC\,2276 one of the richest environments for this still enigmatic bright sources. 

%\subsection{Associations/nature of the ULXs}

We have seen in \citet{wolter11} that six of the X-ray bright sources, four of which are ULXs, are positionally coincident ($1''$--2$''$) with H\textsc{ii} regions \citep[from][]{hodge83,davis97}. However the spatial resolution at the distance of NGC\,2276 is not enough to warrant a physical association. 
At least five supernovae have been observed in the last fifty years (see Section\,\ref{Intro} for references), two close to the center and two in the western region where ULXs are found. 

All observations concur on linking the high activity with the presence of a large number of ULXs, although there is no one-to-one correspondence between peaks in different bands, as is already seen in other galaxies, like the Cartwheel \citep{wolter04}. Radio emission is enhanced throughout the galaxy body: one of the sULX is possibly associated to either a radio bubble or an extended jet structure \citep{mezcua13}, however the complex radio field makes the interpretation difficult. The source has a very hard spectrum: we cannot exclude the possibility that it is a chance superposition with a background AGN.

% \subsection{Number of (BH) ULXs}
We detect 15 sources associated with the galaxy NGC\,2276 in the \cxo\ observation, out of which 8 are above the ULX luminosity threshold. We derive an XLF and confirm the previous estimates of the SFR of NGC\,2276 in the range $\rm SFR = 5$--15~M$_{\sun}$~yr$^{-1}$. The value is close to those of the Antennae and the Cartwheel, which are fitted with SFR = 7.1~M$_{\sun}$~yr$^{-1}$ \citep{zf02} and $\sim 20$~M$_{\sun}$~yr$^{-1}$ \citep{wolter04}, respectively.
These numbers reconcile expectation from \citet{mapelli10} with observations.

%\subsection{Variability of the ULXs in the field of the \xmm\ IXO candidate} 
We know that moderate variability is rather common in ULXs (see e.g. \citealt*{crivellari09} and \citealt{zezas06}) and in recent years a few transient ULXs have also been identified (e.g. \citealt{sivakoff08short,middleton12,middleton13short,soria12,esposito13}). We investigate, as much as possible given the different instruments used with widely different spatial resolution and response, flux variations for the bright source detected in \xmm. 
We do not find any significant evidence for variablity, although a better comparison could be done by using the same instrument repeatedly.

We confirm parameters for the diffuse emission as derived by \citet{rasmussen06}, however we find that the spectrum is best fitted by adding an intrinsic absorption component that raises the unabsorbed luminosity to $L_{\mathrm{X}}^{(0.5-2\,\mathrm{keV})} = 1.8\times 10^{41}$~\lum.
The peak of this distribution is at the center of the galaxy. 
%\subsection{The nuclear source}
The spectrum is consistent with the sum of a hot plasma diffuse component, plus an unresolved component due to active stars and binaries account for a small fraction of the total luminosity (see Section\,\ref{sec:diffuse}). 

The \cxo\ data display no point source at the center of the galaxy. A low luminosity variable AGN of $L_{\mathrm{X}} \leq \mathrm{few} \times 10^{39}$\lum might lurk in the midst.
However the observational results of both the \rst\ HRI \citep[Total $L_{\mathrm{X}} = 3.2\times 10^{41}d^2_{45.7}$\lum\ in a 22$^{\prime\prime}$ radius circle:][]{davis97} and the \xmm\ \citep[$L_{\mathrm{X}} \leq 2.4\times 10^{39}d^2_{45.7}$\lum \,above the diffuse emission:][]{davis04} datasets are consistent with no point source in the center of the galaxy. 
If anything, the presence of two ULX (S11 and S12) at $r\leq 30^{\prime\prime}$  could be responsible for some variability on long time scales.

%\subsection{The forces shaping NGC\,2276}
We have performed a number of simulations to derive the source of activity and asymmetry in NGC\,2276 and confirm that this is linked to the large number of ULXs found. Assuming a simplified scenario we derived a morphology very reminiscent of the observed one and we derive a total mass within a factor of 2 from observations. The presence of a gas envelope around the central elliptical galaxy NGC\,2300 is necessary in order to produce the effects, along the same lines as suggested by \citet{rasmussen06}.

In conclusion, we found that ram-pressure stripping and viscous stripping are the dominant processes and can lead both to the formation of the tails and to the deformation of the front side of the disc galaxy. Tidal forces contribute marginally, by producing tidal arms and thickening the gaseous disc. 
%%%% NOSFR %%%% We found a factor of $2-5$ lower SFR than in the observations, but we are aware that our very approximate treatment of gas underestimates cooling and SFR. 
%%Bearing these {\it caveats} in mind, we conclude that our simulations are able to reproduce the very peculiar shape of NGC\,2276 and the mass of the tidal tails.
Therefore, even if at a simple level, we conclude that our simulations are able to reproduce the very peculiar shape of NGC\,2276 and the mass of the tidal tails.

%sMore precise inital conditions and/or a correct treatment of thermal variations could better reproduce the observed SFR - to match the X-ray derived one.
%Conclusions: we need both tidal interactions and transport phenomena due to the action of the ICM on the disc of NGC\,2276

\section*{Acknowledgments} 
We thank Stefano Rota and Mattia Villani for their contribution in an earlier stage of the analysis. We thank Ewan O'Sullivan for precious suggestions. We thank Dong-Woo Kim, Mar Mezcua and the referees for useful inputs that helped improve the paper.
We acknowledge partial financial support from INAF through grant PRIN-2011-1 ({\it Challenging Ultraluminous X-ray sources: chasing their black holes and formation pathways}). MM and AW acknowledge financial support from the Italian Ministry of Education, University and Research (MIUR) through grant FIRB 2012 RBFR12PM1F. PE acknowledges a Fulbright Research Scholar grant administered by the U.S.--Italy Fulbright Commission and is grateful to the Harvard--Smithsonian Center for Astrophysics for hosting him during his Fulbright exchange. (The Fulbright Scholar Program is sponsored by the U.S. Department of State and administered by CIES, a division of IIE.)
This research is based on data obtained from the \cxo\ Data Archive and has made use of software provided by the \cxo\ X-ray Center (CXC) in the application package \textsc{ciao}. This research has made use of the NASA/IPAC Extragalactic Database (NED) which is operated by the Jet Propulsion Laboratory, California Institute of Technology, under contract with the NASA and of the SIMBAD database, operated at CDS, Strasbourg, France. The Digitized Sky Surveys were produced at the Space Telescope Science Institute under U.S. Government grant NAG W-2166. The simulations were performed with the Lagrange cluster at the Consorzio Interuniversitario Lombardo per L'Elaborazione Automatica (CILEA).

\bibliographystyle{mn2e}
\bibliography{biblio}
\bsp
\label{lastpage}
\end{document}